\documentclass[pra,twocolumn,showpacs,letterpaper,showpacs,superscriptaddress]{revtex4}

\usepackage{graphicx,amsfonts,latexsym,color,dcolumn,bm,epsfig,subfigure}
\usepackage{amssymb}
\usepackage{amsmath}

\renewcommand{\imath}[0]{\mathrm{i}}

\newcommand{\mathbfh}[1]{\hat{\mathbf{#1}}}

\newcommand{\arctanh}[0]{\text{arctanh}}

\usepackage[english]{babel}

\begin{document}

\title{Fluorescence in nonlocal dissipative periodic structures} 

\author{Francesco Intravaia}
\affiliation{Max-Born-Institut, 12489 Berlin, Germany}
\author{Kurt Busch}
\affiliation{Max-Born-Institut, 12489 Berlin, Germany}
\affiliation{Humboldt-Universit\"at zu Berlin, Institut f\"ur Physik, 
             AG Theoretische Optik \& Photonik, 12489 Berlin, Germany}

\date{\today}

\begin{abstract}
We present an approach for the description of fluorescence from optically active material embedded 
in layered periodic structures. Based on an exact electromagnetic Green's tensor analysis, we 
determine the radiative properties of emitters such as the local photonic density of states, Lamb 
shifts, line widths etc. for a finite or infinite sequence of thin alternating plasmonic and dielectric 
layers. In the effective medium limit, these systems may exhibit hyperbolic dispersion relations so 
that the large wave-vector characteristics of all constituents and processes become relevant. These 
include the finite thickness of the layers, the nonlocal properties of the constituent metals, and 
local-field corrections associated with an emitter's dielectric environment. In particular, we show 
that the corresponding effects are non-additive and lead to considerable modifications of an emitter's 
luminescence properties.
\end{abstract}

\maketitle

\section{Introduction}
\label{intro}

Modern technology relies more and more on the ability of building microscopic devices based on 
carefully designed nano-structured materials. For example, engineered stacks of differently doped 
semiconductors are used in modern transistors and (nano-structured) dielectrics with different 
indices of refraction are used to guide light in the most exotic ways. 
A special class of nano-structures consists of alternating metallic and insulating layers arranged 
into one-dimensional periodic lattices. When carefully designed such combinations of plasmonic 
and dielectric materials lead -- within the effective medium limit -- to effective hyperbolic 
dispersion relations 
\cite{Iorsh12,Kidwai12,Poddubny13} 
that may be exploited for a number of applications such as subwavelength imaging 
\cite{Belov06}, 
strong nonlinearities 
\cite{Wurtz11}, 
emission engineering 
\cite{Cortes12}, 
and many more. 
The dispersion relations of these fictitious, spatially uniform hyperbolic meta-materials (HMMs), 
support radiative modes with infinitely large wave vectors which, in turn, lead to broadband 
super-singularities of the local photonic density of states DOS (LDOS). It has been recognized that 
in the actual nano-structure (i.e., without the effective medium description) these singularities 
become regularized 
(i) by the fact that for sufficiently large wave-vector values the actual lattice-structure will be 
resolved 
\cite{Iorsh12} 
and 
(ii) via the nonlocal properties of the metallic constituents 
\cite{Iorsh12,Yan12}.
The LDOS itself can be obtained from the electromagnetic Green's tensor of the relevant structure. 
In turn, the LDOS modifies the spontaneous decay rate of an emitter and the corresponding so-called 
Purcell factor can be determined from the electromagnetic Green's tensor \cite{Yan12}.

Most of the above computations, however, have ignored the effects of local-field corrections in the 
dielectric layers, despite the fact that these corrections, too, may significantly contribute to the 
decay dynamics of the emitter. Therefore, in the present work, we include these effects into the Green's 
tensor formalism and provide a comprehensive study of the non-additive interplay between the above
regularization mechanisms and the local-field corrections on the luminescence properties of emitters 
embedded in finite-sized and infinitely extended layered HMM-type structures. Besides the aforementioned
modified decay rates, this also includes frequency shifts that emitters
experience when being embedded in such structures. In addition, we study these effects in the context
of different material models for the nonlocal optical properties of the plasmonic constituents and 
identify analogies and differences in the results.

\section{Decay Rate and Frequency Shift}
\label{decay-shift}
The dynamical properties of emitters are correlated with the environment surrounding them. In particular, 
the decay rates from excited states or the intrinsic transition frequencies depend on the LDOS associated 
with the electromagnetic environment. Since the latter is strongly correlated with the geometry and the 
optical properties of the objects filling the space around the emitter, it is not surprising that an 
appropriate choice of these two characteristics can lead to enhancement or suppression of decay rates 
as well as to frequency shifts of transitions with respect to their vacuum (empty space) values. Even 
strong non-Markovian effects can be realized
\cite{Hoeppe12}.
In this section we briefly review the general approach for computing decay rates and frequency shifts 
of emitters that are embedded in an arbitrary electromagnetic environment.

\subsection{Spontaneous decay and local field corrections}
We will focus first on the spontaneous decay and calculate the (Purcell) enhancement factor. Within the 
standard theory for a simple two-level emitter with transition frequency $\Omega$ located at 
$\mathbf{r}=\mathbf{r}_{0}$, an emitter's transition rate is given by
\cite{Wylie84,Novotny12}
\begin{equation}
   \Gamma(\Omega) = \mathrm{Tr}\left[
	                                 \frac{2\langle\mathbfh{d}\mathbfh{d}\rangle}{\hbar } 
	                                 \frac{k_{0}^{2}}{\epsilon_{0}}
																	 \mathrm{Im} \left[ \underline{G}(\mathbf{r}_{0},\mathbf{r}_{0},\Omega) \right]
															 \right]
\label{spontan}
\end{equation}
Here, $\epsilon_0$ and $\mathbfh{d}$ denote, respectively, the vacuum permittivity and the emitter's 
dipole operator while $k_{0}=\Omega/c$ represents the free-space wave vector. 
The quantum average $\langle . \rangle$ is  performed over the emitter's ground state. The quantity 
$\underline{G}(\mathbf{r}_{0},\mathbf{r}_{0},\Omega)$ is the electric Green's tensor, i.e., the 
solution to the Maxwell equations for an electric point-dipole oscillating with frequency $\Omega$ 
located at $\mathbf{r}_{0}$ subject to the boundary conditions imposed by the structure surrounding 
the emitter.

In vacuum the spontaneous decay is well defined and has been studied by many authors. In this case 
the Green's tensor is
\begin{align}
	 \underline{G}_{0}(\mathbf{r}_{\alpha},\mathbf{r}_{\beta}; \omega) 
	 &=
   \frac{k_0}{4\pi}\bigg{\{}
       e^{\imath k_0 r_{\alpha,\beta}}
			\bigg{[}
			\frac{3\mathbf{n}_{\alpha,\beta}\mathbf{n}_{\alpha,\beta}-1}{k^{3}_0r_{\alpha,\beta}^{3}}
			\left(1-\imath k_0 r_{\alpha,\beta}\right) \nonumber\\
   &+
	 \frac{1-\mathbf{n}_{\alpha,\beta}\mathbf{n}_{\alpha,\beta}}{k^{3}_0r_{\alpha,\beta}^{3}}
	 (k_0r_{\alpha,\beta})^{2}\bigg{]}
   -\frac{4\pi}{3 k_{0}^{3}}\delta(\mathbf{r}_{\alpha}-\mathbf{r}_{\beta})\bigg{\}}.
\label{homo}
\end{align}
In this expression, we have decomposed the vector 
$\mathbf{r}_{\alpha}-\mathbf{r}_{\beta} = \mathbf{n}_{\alpha,\beta} \, r_{\alpha,\beta}$ into 
a unit vector $\mathbf{n}_{\alpha,\beta}$ and a length $r_{\alpha,\beta}$. Equation \eqref{homo} 
is nothing but the electromagnetic field emitted by a dipole radiating in vacuum 
\cite{Jackson75}.
Using this expression, it is straightforward to show
\begin{align}
   \mathrm{Im} \left[ \underline{G}_{0}(\mathbf{r}_{0},\mathbf{r}_{0},\omega) \right] 
	 &= 
	 \frac{k_{0}}{6\pi} \\
   \Gamma_{0}(\Omega)
	 &=
	 \frac{\langle\langle \hat{d}^{2} \rangle\rangle_{g}}{3\pi\hbar} 
	 \frac{k_{0}^{3}}{\epsilon_{0}}
	 =
	 \frac{c}{2\pi \epsilon_{0}} \alpha_{g}  k_{0}^{4}.
\end{align}
Here, we have also averaged over the dipole's direction so that 
$\langle\hat{d}_{i}\hat{d}_{j}\rangle = \delta_{ij}\langle\langle \hat{d}^{2} \rangle\rangle/3$. 
In the last term we have used the relation 
$\alpha_{g} = \frac{2}{3\hbar \omega_{a}} \langle\langle \hat{d}^{2} \rangle\rangle_{g}$ that
connects the ground-state and angle-averaged dipole moment with the static polarizability 
$\alpha_{g}$. When the dipole is instead embedded in a homogeneous dielectric medium, things are, 
however, more complicated. 
The corresponding Green's tensor can be obtained by the formal replacement 
$k_{0}\to k_{h} = \sqrt{\epsilon(\Omega)}k_{0}$ where $\epsilon(\Omega)$ is the dielectric 
function describing the medium. 
From simple considerations, one would expect that 
$\Gamma(\Omega) = \mathrm{Re}[\sqrt{\epsilon(\Omega)}]\Gamma_{0}(\Omega)$. In reality, in a naive 
application of Eq.\eqref{spontan}, with the Green's tensor given by Eq.\eqref{homo} for a homogeneous 
dielectric medium, even a minute amount of 
\emph{dissipation} 
would lead to a divergent decay rate. Physically, this difficulty can be understood by thinking 
that in a continuum approximation, resulting from a macroscopic average, the emitter can superpose 
with an atom of the dielectric leading to a divergent interaction 
\cite{Fleischhauer99}.
The spontaneous decay of an emitter embedded in a dielectric medium thus represents an example of 
how dissipation can substantially complicate the theory of quantum phenomena. 

\begin{figure}[h]
\includegraphics[width=7cm]{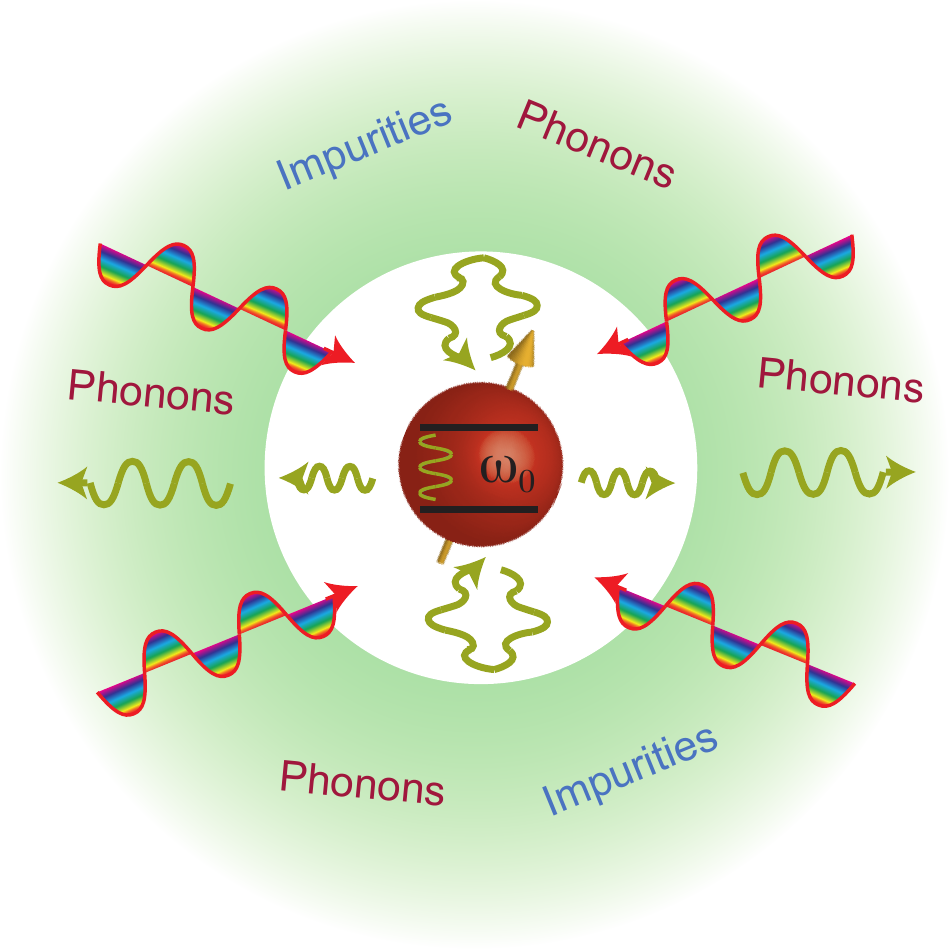}
\caption{(Color online) Within the real-cavity model the emitter is placed in vacuum and a spherical cavity is carved 
         inside the host dielectric material. This is a simple description of realistic situation where 
         the emitter is embedded into the lattice formed by the atoms (molecules) of the host material 
         \cite{Vries98}. 
         The cavity radius is determind by the average distance between the emitter and the nearest 
         atoms (molecules) of the host material.}
\label{fig:real-cavity}
\end{figure}

For the realistic description of spontaneous decay processes we have to take into 
account that the electromagnetic field ``seen'' by an emitter that is embedded in a medium is \emph{not} 
given by solutions of the \emph{macroscopic} Maxwell equations. Rather, the fields provided by 
macroscopic electrodynamics are the result of spatial averages where materials are described 
as continuous entities characterized by permittivities and permeabilityies. At the microscopic 
level the ``granular'' structure of the medium becomes important and the \emph{local field} felt 
by the emitter can be rather different from the result of the macroscopic averaging procedure.
In turn, this may have a very significant impact on the emitters' dynamics, notably if they are 
exposed to multiple scattering effects originating from a complex nano-structured environment. 
In order to treat the local-field problem, one can distinguish two different cases 
\cite{Glauber91,Vries98,Scheel99a,Scheel99,Fleischhauer99}: 
(a) The emitter is of the same species as the atoms (or molecules) that constitute the host medium
or
(b) the emitter is of a different species as the host medium, i.e., an impurity, a substitution etc. 
In both cases (and even for non-dissipative media), the spontaneous decay is non-trivially modified 
with respect to the result for vacuum presented above. The first scenario is described within
the virtual-cavity model \cite{Vries98,Scheel99a,Scheel99,Fleischhauer99}, where the Green's tensor 
is appropriately modified to remove all unphysical divergences. The second scenario is treated within 
the Onsager real-cavity model where the emitter is thought to be placed in vacuum at the center of 
a spherical cavity carved into the host medium \cite{Glauber91,Vries98,Scheel99a,Scheel99,Tomas01,Dung06},
where the cavity radius is essentially given by the distance to the next atom (molecule) of the dielectric
(see Fig. \ref{fig:real-cavity}). 
In the present work, we will restrict ourselves to systems for which the real-cavity model
may be applied.

Before describing how Eq. \eqref{spontan} has to be modified in order to include the local-field 
correction, we first consider the impact of multiple scattering effects by complex nano-structured
systems on the electromagnetic Green's tensor. If all media are linear, we have that
\begin{equation}
   \underline{G}(\mathbf{r}_{0},\mathbf{r}_{0};\omega) 
   = 
   \underline{G}_{h}(\mathbf{r}_{0},\mathbf{r}_{0};\omega) + 
   \underline{G}_{s}(\mathbf{r}_{0},\mathbf{r}_{0}; \omega),
\end{equation}
where the tensor $\underline{G}_{h}$ describes the propagation of the electromagnetic field within a 
homogeneous medium with the same permittivity and permeability as the medium at position $\mathbf{r}_{0}$.
Within the real-cavity model this allows us to identify vacuum as the homogeneous medium 
required for the homogeneous Green's tensor, i.e., we have $\underline{G}_{h} = \underline{G}_0$.
Further, the tensor $\underline{G}_{s}$ represents the scattered Green's tensor and takes into account 
the multiple scattering processes due to the material interfaces in the nano-structure.
For simple geometries the expressions for $\underline{G}_{s}$ are well-known and have been reported 
in the literature (we refer to
\cite{Wylie84,Wylie85,Jackson75,Tomas95}
and our discussion of layered systems below). 
In the local-field corrected Green's tensor within the framework of the real-cavity model, the 
emitter is placed in vacuum enclosed at the center of a  spherical cavity with radius $R$ much smaller 
than the optical wavelength ($k_{0}R\ll1$). In our 
specific case of alternating metallic and dielectric layers, this spherical local-field cavity with the 
emitter in its center is located inside a central dielectric layer with permittivity $\epsilon(\omega)$ 
sandwiched between a finite or infinite number of further layers on either side. 
This means that the scattering Green's tensor $\underline{G}_{s}$ is determined from both, the boundary 
conditions on the cavity sphere and on interfaces between the layers. Clearly, these two sets of boundary 
conditions act in a very non-additive way, thus complicating the evaluation of the Green's tensor. 
Fortunately, this topic has been extensively discussed in the literature 
\cite{Tomas01,Dung06,Sambale07,Scheel08} 
so that we may directly utilize that in this case the scattered Green's tensor can be written as
\begin{equation}
   \underline{G}_{s}(\mathbf{r}_{0},\mathbf{r}_{0}; \omega)
   =
   C(\omega,R)\frac{k_{0}}{6\pi} + 
   S^{2}(\omega,R) \, \underline{\mathcal{G}}_{s}(\mathbf{r}_{0},\mathbf{r}_{0}; \omega)
\label{CorrectedGs}
\end{equation}
Here,  $\underline{\mathcal{G}}_{s}$ is the scattered Green's tensor that exclusively results from the 
multiple scattering at the layers' interfaces and the effect of the spherical cavity appears in two terms: 
An offset $C(\omega,R)$ that will survive even if we remove all the scattering from all layers and a 
prefactor $S^2(\omega,R)$ to the scattering Green's tensor $\underline{\mathcal{G}}_{s}$ that account 
for the multiple scattering at layer interfaces. This specific 
structure of the full scattering Green's tensor $\underline{G}_{s}$ makes quite explicit the non-additive 
character of the local-field and the multiple-scattering corrections. Within the real-cavity model, we have  
\cite{Tomas01,Dung06,Sambale07,Scheel08}
\begin{align}
   C(\omega,R) 
   = 
   & \frac{h^{(1)}_{1}(\nu_{0})[\nu h^{(1)}_{1}(\nu)] -  
           \epsilon(\omega)[h^{(1)}_{1}(\nu_{0})\nu_{0}]'h^{(1)}_{1}(\nu)}{\epsilon(\omega)h^{(1)}_{1}(\nu)[\nu_{0} j_{1}(\nu_{0})]'
                                                                           -[h^{(1)}_{1}(\nu)\nu]'j_{1}(\nu_{0})} \\
   S(\omega,R)
   =
   & \frac{j_{1}(\nu_{0})[\nu_{0}h^{(1)}_{1}(\nu_{0})]' -
           [j_{1}(\nu_{0})\nu_{0}]'h^{(1)}_{1}(\nu_{0})}{j_{1}(\nu_{0})[\nu h^{(1)}_{1}(\nu)]'
                                                         -\epsilon(\omega)[j_{1}(\nu_{0})\nu_{0}]'h^{(1)}_{1}(\nu)}
\end{align}
where we have introduced the abbreviations $\nu_{0}=k_{0} R$ and $\nu=k_{0} \sqrt{\epsilon(\omega)} R$ while
$j_{1}(\nu)$ and $h^{(1)}_{1}(\nu)$ denote, respectively, the spherical Bessel function of order one and the 
spherical Hankel function of the first kind of order one (the prime indicates the derivative with 
respect to the argument of the corresponding Bessel/Hankel functions). We would like to emphasize 
that $\nu$ and, consequently, also the wave vector of the dielectric host medium $ k_h = k_{0} \sqrt{\epsilon(\omega)}$ 
are complex valued due to the fact that the host medium's permittivity $\epsilon(\omega)$ is, in general, 
complex valued.
Also, as a result of the foregoing, the spontaneous decay formally depends on an external parameter, i.e., the cavity 
radius $R$. As stated above, this parameter is fixed by the microscopic arrangement of the host dielectric
atoms and must be experimentally determined. However, for optical frequencies, one may even consider an expansion 
of the above expressions in the small parameter $k_0R \ll 1$(see Refs. \cite{Tomas01,Dung06,Sambale07,Scheel08}).
 
\begin{figure}[hhh]
\includegraphics[width=8cm]{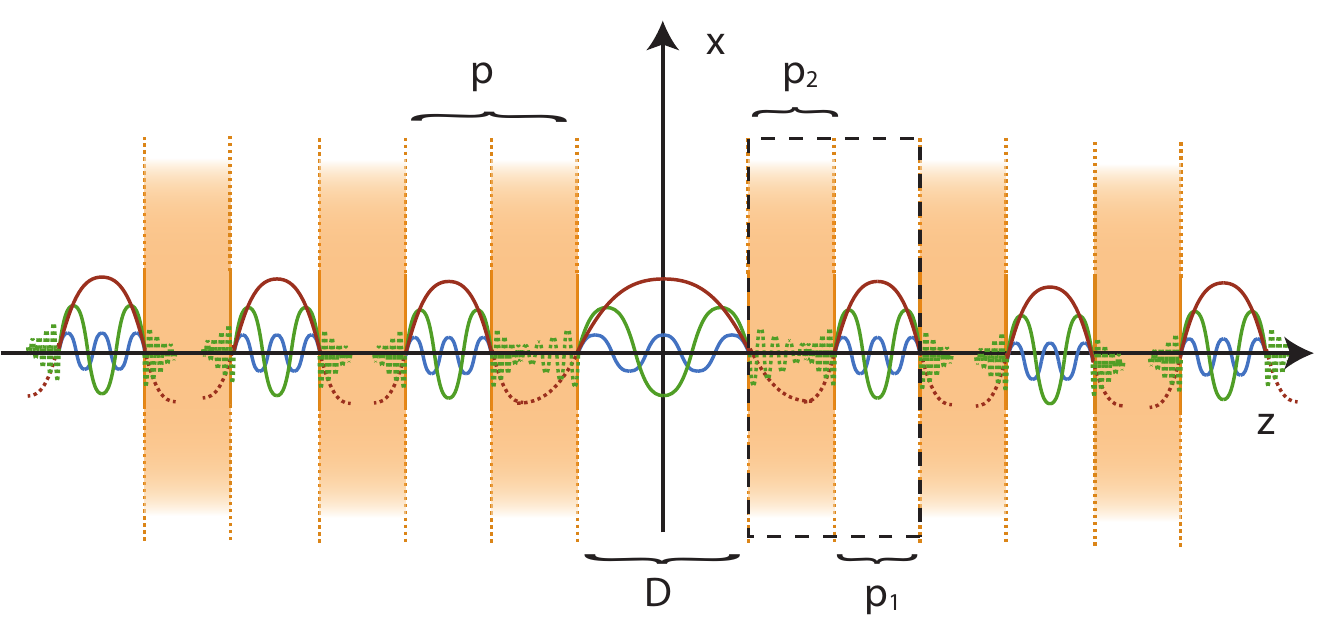} 
\caption{(Color online) Sketch of the layered material considered in this work. A central dielectric layer (thickness $D$,
         white shading) is sandwiched between a finite or infinite number of identical bilayers (total 
         thickness $p = p_1 + p_2$) consisting of a plasmonic (thickness $p_2$, red shading) and a dielectric 
         layer (thickness $p_1$, white shading). All dielectric layers are made from the same material 
         (permittivity $\epsilon(\omega)$). The plasmonic layers are identical and different material 
         models are considered. For the infinite system, the dashed region delineates the unit cell.
         See text for further details.}
\label{ModulatedRegionHH}
\end{figure}

Upon averaging over the emitter's dipole orientations, the local-field corrected emission rate is given by
\begin{align}
    \Gamma(\Omega)
    &=  \frac{1}{3} \, \mathrm{Tr}[\underline{H}_{\rm P}] \,  \Gamma_{0}(\Omega) ,
\end{align}
where we have introduced the tensor
\begin{align}
   \underline{H}_{\rm P}
   &=
   1 + \mathrm{Im}[C(\Omega,R)] \nonumber \\
     &+ \frac{6\pi}{k_{0}} \mathrm{Im} \left[S^{2}(\Omega,R) \,
                                            \underline{\mathcal{G}}_{s}(\mathbf{r}_{0},\mathbf{r}_{0}; \Omega)
                                      \right].
\end{align}
The last step in our approach is the determination of $\underline{\mathcal{G}}_{s}$, for which we 
have to specify the arrangement of layers. In this work, we consider one central dielectric layer 
containing the emitter that is symmetrically sandwiched by a finite or infinite number of identical 
bilayers that consist of one plasmonic and one dielectric material. For simplicity, we restrict 
ourselves to the case where all dielectric layers are made from the same material 
(see Fig. \ref{ModulatedRegionHH}).
From the emitter's point of view, the structure resembles a cavity that is formed by two Bragg mirrors 
and is filled with a dielectric material with permittivity $\epsilon(\omega)$. We align the $z$-direction 
with the stacking direction of the layers and denote the emitter's distance from the nearest plasmonic 
layer with $d$. Further, the left and right Bragg mirror are, respectively, labeled with indices ``1'' 
and ``2''. Then, we may decompose the scattering Green's tensor $\underline{\mathcal{G}}_{s}$ into 
components parallel and perpendicular to the stacking direction of the layers 
\begin{align}
   \underline{\mathcal{G}}_{s}(\mathbf{r}_{0},\mathbf{r}_{0};\omega)
   &\equiv
   \underline{\mathcal{G}}_{s}(d,\omega) \nonumber\\
   &= 
   \mathcal{G}_{\|}(d,\omega)(\mathbf{x}\mathbf{x}+\mathbf{y}\mathbf{y}) + 
   \mathcal{G}_{\bot}(d,\omega)\mathbf{z}\mathbf{z}.
\end{align} 
In turn, the parallel and perpendicular components of the Green's tensor, $\mathcal{G}_{\|}$ and $\mathcal{G}_{\bot}$,
may be expressed through the structure's geometrical parameters and the (frequency- and wave-vector 
dependent) reflection coefficients $r^{\sigma}_1$, and $r^{\sigma}_2$ ($\sigma=s,p$) of the two Bragg 
mirrors for s- and p-polarized plane waves. Specifically, if $D$ is the cavity length, these components 
of the Green's tensor read as \cite{Tomas95}
\begin{widetext}
\begin{subequations}
\begin{align}
   \mathcal{G}_{\|}(d,\omega) 
   = 
   \frac{1}{8\pi} \int_{0}^{\infty}dk\, \frac{k \,\kappa_{\rm lc}}{k_{h}^{2}}
   &\left[ \frac{r^{p}_{1}e^{-2\kappa_{\rm lc} d}+r^{p}_{2}e^{-2\kappa_{\rm lc}(D-d)}-2 r^{p}_{1}r^{p}_{2}e^{-2\kappa_{\rm lc}D}}{1-r^{p}_{1}r^{p}_{2}
                                                                                                                                  e^{-2\kappa_{\rm lc} D}} \right. \nonumber\\
  +&\left. \frac{k_{h}^{2}}{\kappa_{\rm lc}^{2}}
           \frac{r^{s}_{1}e^{-2\kappa_{\rm lc} d}+r^{s}_{2}e^{-2\kappa_{\rm lc}(D-d)}+2 r^{s}_{1}r^{s}_{2}e^{-2\kappa_{\rm lc} D}}{1-r^{s}_{1}r^{s}_{2}
                                                                                                                                   e^{-2\kappa_{\rm lc} D}}
   \right],
\end{align}
\begin{equation}
   \mathcal{G}_{\bot}(d,\omega)
   =
   \frac{1}{8\pi} \int_{0}^{\infty}dk\, \frac{k \,\kappa_{\rm lc}}{k_{h}^{2}} \,
   \left[2\frac{k^{2}}{\kappa_{\rm lc}^{2}} 
          \frac{r^{p}_{1}e^{-2\kappa_{\rm lc} d}+r^{p}_{2}e^{-2\kappa_{\rm lc}(D-d)}+2 r^{p}_{1}r^{p}_{2}e^{-2\kappa_{\rm lc} D}}{1-r^{p}_{1}r^{p}_{2}
                                                                                                                                  e^{-2\kappa_{\rm lc}D}}
   \right],
\end{equation}
\label{greenT}
\end{subequations}
\end{widetext}
where $\kappa_{\rm lc}=\sqrt{k^{2}-k_{h}^{2}}=-\imath k_{\rm lc}$. For our symmetric geometry, the reflection 
coefficients of the Bragg mirrors are identical so that $r^s_{1} = r^s_{2} = r^s$ and $r^p_{1}=r^p_{2} = r^p$. 
For simplicity, we will subsequently restrict ourselves to the case of $D=2 d$ so that the emitter is located 
at the center of the cavity (see Fig. \ref{ModulatedRegionHH}). The complete determination of the scattered 
Green's tensor components requires the evaluation of the reflection coefficient 
$r^{\sigma}$ which will be the subject of the section \ref{sec:coefficients}.

\subsection{Frequency shift}
In addition to the spontaneous decay rate, modifications of the electromagnetic environment as well as local-field 
corrections have an impact on the emitter's energy levels. The treatment of this effect leverages on the same tools as
discussed above, i.e., the determination of the Green's tensor. For a two-level emitter the standard approach 
relies on the connection between the Casimir-Polder energy ($U_{\rm CP}$) \cite{Intravaia11,Haakh13} and the change 
in the emitter's ground state energy ($\Delta E_{g}$)
\cite{Wylie84,Wylie85}: 
$U_{\rm CP}=\Delta E_{g}\equiv \hbar \Delta \omega_{g}$. (This connection can also be generalized to higher energy 
levels 
\cite{Wylie84,Wylie85}.) 
If $\underline{\alpha}(\omega)$ is the ground-state polarizability, the Casimir-Polder energy is given by
\begin{equation}
   U_{\rm CP} = 
   \hbar\int_{0}^{\infty}\frac{d\xi}{2\pi\epsilon_0} 
   \mathrm{Tr}\left[\underline{\alpha}(\imath \xi)\cdot \frac{\xi^2}{c^2}\underline{G}_{s}(\mathbf{r}_{0},\mathbf{r}_{0}; \imath\xi)\right],
\end{equation}
where we have used the local-field corrected expression \eqref{CorrectedGs} for the scattered Green's tensor. For a 
two-level emitter and after averaging over the emitter's dipole direction, we obtain
\begin{equation}
   \underline{\alpha}(\imath \xi) = \alpha(\imath \xi) \underline{1} 
                                  = \alpha_{g}\frac{\Omega^{2}}{\xi^{2}+\Omega^{2}} \underline{1} \, ,
\end{equation}
where $\underline{1}$ denotes the unit tensor. Within the framework of second-order perturbation theory  
\cite{Wylie84,Wylie85,Intravaia11} 
the above expression is equivalent to an additional contribution to the vacuum-induced Lamb shift of the 
ground state energy that is generated by the nano-structure. From Eq.\eqref{CorrectedGs}, we obtain two 
distinct contributions. The first is exclusively associated with the local-field correction 
\begin{equation}
   U^{(1)}_{\rm CP} = 3\hbar\int_{0}^{\infty}\frac{d\xi}{2\pi \epsilon_0} \, \alpha(\imath \xi) \frac{\xi^2}{c^2} G_{\rm C} (\imath\xi,R),
\end{equation}
where
\begin{equation}
   G_{\rm C} (\omega,R) = C(\omega,R)\frac{k_{0}}{6\pi}.
\end{equation}
The second contribution
\begin{equation}
   U^{(2)}_{\rm CP} 
   = \hbar \int_{0}^{\infty}\frac{d\xi}{2\pi \epsilon_0} \, \alpha(\imath \xi) \,
                                               S^2(\imath \xi, R)
                                               \mathrm{Tr}\left[\underline{\mathcal{G}}_{s}(d, \imath\xi)\right],
\label{geoshift}
\end{equation}
describes the impact of the Bragg mirrors on the shift and, therefore, depends on the detailed characteristics 
of the structure such as the distance of the emitter from the dielectric/metal interfaces etc. In the following,
we will exclusively focus on this second contribution, since for a specific dielectric it is 
the only one that can be tuned as a function on the geometric parameters of the surrounding nano-structure \cite{Sambale07}.

\section{Periodic Structures: Bloch equation and Scattering Coefficients}
\label{sec:coefficients}
The behavior of the electromagnetic field within a periodic structure can be analyzed in the framework of 
the Bloch theorem 
\cite{Busch07}.
If the periodic medium is composed of a stacking sequence of layers, the problem can be tackled analytically 
with the help of a transfer-matrix approach
\cite{Yeh77,Yariv83}. 
Specifically, the continuity of the transverse field components leads to boundary conditions across each 
layer. For the metallic layer, we have (c.f. Fig. \ref{ModulatedRegionHH})
\begin{equation}
   \begin{pmatrix}
      E \\
      c B
   \end{pmatrix}_{z=\frac{p_{1}}{2}+p_{2}}^{\sigma}
   =
   \mathbb{M}^{\sigma}_{\rm nl}(p_{2})
   \begin{pmatrix}
      E \\
      c B
   \end{pmatrix}_{z=\frac{p_{1}}{2}}^{\sigma},
\end{equation}
while for the  dielectric layer, we obtain
\begin{equation}
   \begin{pmatrix}
      E \\
      c B
   \end{pmatrix}_{z=\frac{p_{1}}{2}+p}^{\sigma}
   =
   \mathbb{M}^{\sigma}_{\rm lc}(p_{1})
   \begin{pmatrix}
      E \\
      c B
   \end{pmatrix}_{z=\frac{p_{1}}{2}+p_{2}}^{\sigma}.
\end{equation}
Here, $\sigma=s,p$ indicates again the polarization while 
$\mathbb{M}^{\sigma}_{\rm nl}\left(p_{2}\right)$ and $\mathbb{M}^{\sigma}_{\rm lc}\left(p_{1}\right)$ 
represent the transfer matrices associated with each layer. Specifically, as we will exclusively consider 
spatially local material models for the dielectric, we have introduced the subscript ``lc'' for dielectric 
layer transfer matrices. Similarly, as we will mainly focus on spatially nonlocal material models for the 
metal, we have introduced the subscript ``nl'' for metal layer transfer matrices. Further, the propagation 
across a bilayer can be written as
\begin{equation}
   \begin{pmatrix}
      E \\
      c B
    \end{pmatrix}_{z=\frac{p_{1}}{2}+p}^{\sigma}
    =
    \mathbb{T}^{\sigma}
    \begin{pmatrix}
       E \\
       c B
    \end{pmatrix}_{z=\frac{p_{1}}{2}}^{\sigma},
\end{equation}
where the corresponding transfer matrix $\mathbb{T}$ is given by
\begin{align}
   \mathbb{T}^{\sigma}
   =
   \mathbb{M}^{\sigma}_{\rm nl}\left(p_{2}\right)\mathbb{M}^{\sigma}_{\rm lc}\left(p_{1}\right).
\end{align}
For a local dielectric material, the transfer matrix $\mathbb{M}^{\sigma}_{\rm lc}\left(p_{1}\right)$ can be 
written as
\begin{equation}
   \mathbb{M}^{\sigma}_{\rm lc}\left(p_{1}\right)
   =
   \begin{pmatrix}
      \cos(k_{\rm lc}p_{1})                                              & \imath \delta^{\sigma}\sin(k_{\rm lc}p_{1})Z^{\sigma}_{\rm lc}\\
      \imath \delta^{\sigma} \sin(k_{\rm lc}p_{1}) / Z^{\sigma}_{\rm lc} & \cos(k_{\rm lc}p_{1})
   \end{pmatrix},
\end{equation}
where, for p- and s-polarization, respectively, we have $\delta^{p}=1$ and $\delta^{s}=-1$. Further, we have 
defined the local surface impedances
\begin{equation}
   Z_{\rm lc}^{p} 
   = 
   \frac{c k_{\rm lc}}{\omega\epsilon(\omega)}, \quad \mbox{and} 
   \quad Z_{\rm lc}^{s}
   =
   \frac{\omega}{c k_{\rm lc}}.
\end{equation}
In the case of a metallic layer with nonlocal material model the entries of $\mathbb{M}_{\rm nl}(p_{2})$ depends 
on the specific model used to describe the nonlocality (see section \ref{nonlocal}). Nevertheless, the matrix  
has a similar structure as in the local case
\begin{equation}
   \mathbb{M}^{\sigma}_{\rm nl}\left(p_{2}\right)
   =
   \begin{pmatrix}
      \mathrm{cs}^{\sigma}_{\rm nl}(p_{2})                           & \imath \delta^{\sigma} \mathcal{Z}^{\sigma}_{\rm right}(p_{2})\\
      \imath \delta^{\sigma} / \mathcal{Z}^{\sigma}_{\rm left}(p_{2}) & \mathrm{cs}_{\rm nl}(p_{2}),
   \end{pmatrix}
   \label{MnlMatrix}
\end{equation}
with functions $\mathrm{cs}^{\sigma}_{\rm nl}(p_{2})$, $\mathcal{Z}^{\sigma}_{\rm left}(p_{2})$ and
$\mathcal{Z}^{\sigma}_{\rm left}(p_{2})$ that have to be determined from the specific material model.
For instance, reciprocity stipulates that both matrices $\mathbb{M}^{\sigma}_{\rm nl}\left(p_{2}\right)$ 
and $\mathbb{M}^{\sigma}_{\rm lc}\left(p_{1}\right)$ have a determinant equal to one. While this clearly 
is fulfilled for $\mathbb{M}_{\rm lc}(p_{1})$, this condition imposes certain restriction to the general 
form describing the non-local case
\begin{equation}
   \frac{\mathcal{Z}_{\rm right}(p_{2})}{\mathcal{Z}_{\rm left}(p_{2})}+\mathrm{cs}^{2}_{\rm nl}(p_{2}) = 1.
\end{equation}

\subsection{Infinite Bragg Mirrors}
In an infinite periodic structure the Bloch theorem further requires that the field values across 
a unit cell satisfy 
\begin{equation}
   \begin{pmatrix}
      E \\
      c B
   \end{pmatrix}_{z=\frac{p_{1}}{2}+p}=e^{\imath \alpha p}
   \begin{pmatrix}
      E \\
      c B
   \end{pmatrix}_{z=\frac{p_{1}}{2}}~.
\end{equation}
This leads to an eigenvalue equation $\mathbb{T}=e^{\imath \alpha p}$ that connects the eigenvalues 
with the generally complex values of the Bloch vector $\alpha$. Since $\mathrm{Det}[\mathbb{T}]=1$, 
the eigenvalues have the form $e^{\pm\imath \alpha p}$ and using the invariance of the trace of a 
matrix we obtain the generalized Bloch equation
\begin{multline}
   \cos(\alpha p) 
   = \cos(k_{\rm lc}p_{1}) \mathrm{cs}_{\rm nl}(p_{2})\\
     - \frac{1}{2}\sin(k_{\rm lc}p_{1})
       \left(\frac{Z_{\rm lc}}{\mathcal{Z}_{\rm left}(p_{2})} + \frac{\mathcal{Z}_{\rm right}(p_{2})}{Z_{\rm lc}}\right),
\end{multline}
where, for simplicity of notation, we have dropped the polarization superscript. 
For the computation of the scattering Green's tensor, we require the reflection coefficients from 
infinitely extended half-spaces and this can be derived in terms of the eigenvector of the transfer matrix 
\cite{Mochan87,Yariv83}: 
Each of the two eigenvectors corresponds to a Bloch mode propagating either to the right or to the left of 
the unit cell. The components of the mode are the corresponding electric and magnetic fields from which it 
is possible to derive the surface impedance $Z_{\rm per}$ for the periodic structure \cite{Mochan87}. We obtain
\begin{equation}
   Z_{\rm per} = -\frac{\mathbb{T}_{12}}{\mathbb{T}_{11}-e^{\imath \alpha p}}
               = -\frac{\mathbb{T}_{22}-e^{\imath \alpha p}}{\mathbb{T}_{21}},
\end{equation}
which, upon inserting the specific form of the transfer matrices, explicitly reads as
\begin{align}
   Z_{\rm per} 
   &= 
   \imath\mathcal{Z}_{\rm right} 
   \frac{\cos(k_{\rm lc}p_{1}) + 
         \frac{Z_{\rm lc}}{\mathcal{Z}_{\rm left}} \mathrm{cs}_{\rm nl}(p_{2}) \sin(k_{\rm lc}p_{1})}{e^{\imath \alpha p} - 
                                                                                                      \cos(k_{\rm lc}p_{1}) \mathrm{cs}_{\rm nl}(p_{2}) +
                                                                                                      \sin(k_{\rm lc}p_{1}) \frac{\mathcal{Z}_{\rm right}}{Z_{\rm lc}}}
                                                                                                      \nonumber\\
   &=
   - \imath \mathcal{Z}_{\rm left}
   \frac{e^{\imath \alpha p} - \cos(k_{\rm lc}p_{1}) \mathrm{cs}_{\rm nl}(p_{2})
         + \sin(k_{\rm lc}p_{1})\frac{Z_{\rm lc}}{\mathcal{Z}_{\rm left}}}{\cos(k_{\rm lc}p_{1}) + 
                                                                           \frac{\mathcal{Z}_{\rm left}}{Z_{\rm lc}} \mathrm{cs}_{\rm nl}(p_{2}) \sin(k_{\rm lc}p_{1})},
\end{align}
where, again, the polarization superscript has been suppressed. The reflection coefficients for infinite
periodic Bragg mirrors can then be 
written as
\begin{equation}
   r^{p} = \frac{Z^{p}_{\rm lc}-Z^{p}_{\rm per}}{Z^{p}_{\rm lc}+Z^{p}_{\rm per}},
   \quad \mbox{and} \quad
   r^{s} = -\frac{Z^{s}_{\rm lc}-Z^{s}_{\rm per}}{Z^{s}_{\rm lc}+Z^{s}_{\rm per}}.
\end{equation}

\subsection{Finite Bragg Mirrors}
The transfer matrix $\mathbb{T}$ can also be utilized for obtaining the reflection coefficients for structures 
with a finite number of bilayers \cite{Yeh77,Yariv83} embedded into half-spaces of the host dielectric materials. 
For a single metallic layer (i.e., a slab) with thickness $p_{2}$ the reflection and transmission coefficients, 
$r_{\rm slab}^{\sigma}$ and $t_{\rm slab}^{\sigma}$, are given by
\begin{equation}
   r_{\rm slab}^{\sigma}
   =
   \delta^{\sigma} 
   \frac{Z^{2}_{\rm lc}-Z_{\rm right}(p_{2}) Z_{\rm left}(p_{2}) }{Z^{2}_{\rm lc} + 
                                                                   Z_{\rm right}(p_{2}) Z_{\rm left}(p_{2}) + 
                                                                   2 \imath\, \mathrm{cs}_{\rm nl}(p_{2}) Z_{\rm lc}Z_{\rm left}}
\end{equation}
\begin{equation}
   t_{\rm slab}^{\sigma}
   =
   \frac{2\imath Z_{\rm lc} Z_{\rm left}(p_{2})}{Z_{\rm lc}^{2} + 
                                                 Z_{\rm right}(p_{2}) Z_{\rm left}(p_{2}) +
                                                 2\imath\, \mathrm{cs}_{\rm nl}(p_{2}) Z_{\rm lc} Z_{\rm left}(p_{2})}.
\end{equation}
Here, again, the polarization superscript has been suppressed.  
The expression for the reflection coefficients of a finite structure composed of a sequence of $N$ can then be 
written as \cite{Yeh77,Yariv83}
\begin{equation}
   r^{\sigma}_{N}
   = 
   \frac{r^{\sigma}_{\rm slab} e^{2\imath k_{\rm lc} p_{1}}}{1 - t^{\sigma}_{\rm slab}e^{\imath k_{\rm lc} p_{1}} 
                                                                \frac{\sin([N-1]\alpha p)}{\sin(N \alpha p)}}, 
\end{equation}
where, we recall that $\alpha$ is the Bloch vector.

\section{The Description of Nonlocal Media}
\label{nonlocal}
The description of the nonlocal properties of metals has been the subject of many publications in the past (we 
refer to \cite{Ford84} for an overview of earlier works) and, in the context of nano-plasmonics, has recently 
experienced renewed interests. Here, we give a brief review of two different models which have been formulated 
in the literature and provide the results that are required for our computations.

\subsection{The SCIB Model}
We first consider the approach which goes under the name of \emph{semi-classical infinite barrier} (SCIB) model 
\cite{Feibelman82,Ford84}, 
that has been utilized for the description of the anomalous skin effect 
\cite{Kliewer68,Jones69,Fuchs69}. 
In the SCIB model, the interface is described very crudely through an infinite barrier but it takes into account
the most relevant physical phenomena inside the metal 
\cite{Feibelman82,Ford84}.
Within this approach, the electrons are treated as a classical ideal gas that is governed by the Fermi-Dirac 
statistics and whose dynamics is described via the Boltzmann equation. It is further assumed that the electrons 
in this nonlocal medium specularly reflect at the interface with another medium
\cite{Horovitz12}. 
It has been shown that the fields in the interior of such a nonlocal medium, are identical to the fields that 
arise from a current sheet source at the surface. Since in our case, the system is invariant with respect to 
translations in the $x-y$ plane and an interface is located at $z=z_{0}$ this sheet current has the form (we 
follow the notation of Ref. \cite{Ford84})
\begin{equation}
   \mathbf{j}(\mathbf{r},t) 
   = 
   \mathbf{J} \delta(z-z_{0}) e^{\imath(\mathbf{k}\cdot\mathbf{R}-\omega t)} 
   \quad \text{ with } \quad 
   \mathbf{J}\cdot\mathbf{z}=0,
\end{equation}
where $\mathbf{k}$ is the component of the wavevector orthogonal to the $z$-direction ($\mathbf{z}$ denotes the 
unit vector along the $z$-direction). The corresponding electric field has the form
\begin{equation}
   \mathbf{E}(\mathbf{r},t) = \mathbf{E}(z) e^{\imath(\mathbf{k}\cdot\mathbf{R}-\omega t)}
\end{equation}
and an analoguous expression holds for the magnetic field. 
If $\mathbf{K}=(\mathbf{k},k_{z})$ is the three-dimensional wave vector, the dielectric tensor of the metal within
SCIB can be written as
\begin{equation}
   \underline{\epsilon}(K,\omega) = \epsilon_{l}(K,\omega)\frac{\mathbf{K}\mathbf{K}}{K^{2}} + 
                                    \epsilon_{t}(K,\omega)\frac{K^{2}\underline{1} -\mathbf{K}\mathbf{K}}{K^{2}}.
\end{equation}
The Maxwell equations can then be solved in terms of the longitudinal and transverse dielectric functions,
$\epsilon_{l}(K,\omega)$ and $\epsilon_{t}(K,\omega)$, respectively, and we obtain that the component of the 
electric and magnetic field parallel to the layer interface can be written as 
\begin{align}
   E_{nl}^{p}(z)
   &=   
   -\frac{Z_{0}}{2}\mathbfh{k}\cdot\mathbf{J} \eta^{p}(z-z_{0}) \\
   B_{nl}^{p}(z)
   &=
   -\frac{Z_{0}}{2c}\mathbfh{k}\cdot\mathbf{J}\beta(z-z_{0}) \\
   E_{nl}^{s}(z)
   &=
   -\frac{Z_{0}}{2}(\mathbfh{z}\times\mathbfh{k})\cdot\mathbf{J}\eta^{s}(z-z_{0}) \\
   B_{nl}^{s}(z)
   &=
   \frac{Z_{0}}{2c}(\mathbfh{z}\times\mathbfh{k})\cdot\mathbf{J}\beta(z-z_{0})
\end{align}
where $Z_{0}=\sqrt{\mu_{0}/\epsilon_{0}}$ is the vacuum impedance and the dimensionless functions $\eta(z)$ and 
$\beta(z)$ are defined as
\begin{subequations}
\begin{equation}
   \eta^{p}(z)
   =
   \frac{\omega}{c}\int_{-\infty}^{\infty}\frac{dk_{z}}{2\pi \imath}
   \left(\frac{2\frac{k_{z}^{2}}{K^{2}}e^{\imath k_{z}z}}{K^{2}-\frac{\omega^{2}}{c^{2}}\epsilon_{t}(K,\omega)} -
         \frac{2\frac{k^{2}}{K^{2}}e^{\imath k_{z}z}}{\frac{\omega^{2}}{c^{2}}\epsilon_{l}(K,\omega)}\right),
\end{equation}
\begin{equation}
   \eta^{s}(z)
   =
   \frac{\omega}{c} \int_{-\infty}^{\infty}\frac{dk_{z}}{2\pi \imath}
   \frac{2 e^{\imath k_{z}z}}{K^{2}-\frac{\omega^{2}}{c^{2}}\epsilon_{t}(K,\omega)},
\end{equation}
\begin{equation}
   \beta(z)
   =
   \int_{-\infty}^{\infty}\frac{dk_{z}}{2\pi \imath} 
   \frac{2 k_{z}e^{\imath k_{z}z}}{K^{2}-\frac{\omega^{2}}{c^{2}}\epsilon_{t}(K,\omega)}.
\end{equation}
\label{etabeta}
\end{subequations}
At this point, we would like to note that $\eta(z)=\eta(-z)$ and $\beta(z)=-\beta(-z)$. 
Also, it is straightforward to show that $\beta(0^{+})= \lim_{z \to 0, z > 0} [\beta(z)] = 1$.

The above expressions still do not provide an essential piece of information, i.e. the expressions 
for the dielectric function. In fact, this is where the Boltzmann equation for the dynamics of the 
electronic fluid (semi-classical approach) has to be employed. In the Boltzmann equation approach, 
the most important aspect is the treatment of collisions. In the single relaxation-time approximation 
the corresponding analysis delivers 
\cite{Ford84,Lindhard54,Kliewer68,Jones69} 
\begin{subequations}
\begin{align}
   \epsilon_{t}(\omega)
   &=
   \epsilon_{b}(\omega) - \frac{\omega^{2}_{p}}{\omega(\omega+\imath \Gamma)}f_{t}(v),\\
   \epsilon_{l}(K,\omega)
   &=
   \epsilon_{b}(\omega)-\frac{\omega^{2}_{p}}{\omega(\omega+\imath \Gamma)}f_{l}(v).
\end{align}
\label{epsilonSCIB}
\end{subequations}
Here, we have introduced the functions \cite{Kliewer68}
\begin{subequations}
\begin{align}
   f_{t}(v)
   &= 
   \frac{3}{2 v^{3}}\left[v-(1-v^{2})\arctanh (v)\right], \\
   f_{l}(v)
   &=
   -\frac{3}{v^{2}}\frac{\omega}{\imath \Gamma}\frac{v-\arctanh (v)}{v\left(1+\frac{\omega}{\imath \Gamma}\right)-\arctanh (v)}.
\end{align}
\label{boltzmann}
\end{subequations}
Further, we have abbreviated $v=v_{F}K/(\omega+\imath \Gamma)$ where $\omega_{p}$, $v_{F}$, and $\Gamma$
denote, respectively, the plasma frequency, the Fermi velocity, and the dissipation rate of the metal. 
In addition, the function $\epsilon_{b}(\omega)$ describes the dielectric function associated with a
potential dynamic behavior of the ionic background charge. In the following, we will disregard this 
contribution for all material models and, consequently, set $\epsilon_b(\omega) \equiv 1$. 
Using the above expressions, we may determine the entries of the transfer matrix 
$\mathbb{M}^{\sigma}_{\rm nl}\left(p_{2}\right)$ and obtain
\begin{subequations}
   \begin{equation}
   \mathrm{cs}_{\rm nl}(p_{2})
   =
   \frac{\beta(0^{+})\eta(0)+\beta(p_{2})\eta(p_{2})}{\beta(p_{2})\eta(0)+\beta(0^{+})\eta(p_{2})}
\end{equation}
\begin{equation}
   \mathcal{Z}_{\rm up}(p_{2})
   = 
   -\imath \frac{\eta^{2}(p_{2})-\eta^{2}(0)}{\beta(p_{2})\eta(0)+\beta(0^{+})\eta(p_{2})}
\end{equation}
\begin{equation}
   \mathcal{Z}_{\rm dwn}(p_{2})
   = 
   \imath \frac{\beta(p_{2})\eta(0)+\beta(0^{+})\eta(p_{2})}{\beta^{2}(p_{2})-\beta^{2}(0^{+})}
\end{equation}
\end{subequations}

\subsection{The Hydrodynamic Model}
The SCIB is just one of the possible approaches for treating the nonlocal behavior of metals and 
alternative, however less realistic \cite{Feibelman82}, descriptions may be based an hydrodynamic 
models. 

The standard approach (for a recent example see, e.g., ref. \cite{raza11}) describes the metal's 
conduction electrons as a compressible fluid and leads to the following equation for the free current 
in the metal
\begin{equation}
   \beta_{\rm H}^{2}\nabla \left( \nabla\cdot \mathbf{j}(\mathbf{r},\omega) \right) +
   \omega(\omega+\imath \Gamma)\mathbf{j}(\mathbf{r},\omega)
   =
   \imath \omega \omega_{p}^{2}\epsilon_{0}\mathbf{E}(\mathbf{r},\omega).
\end{equation}
Here, $\beta_{\rm H}$ describes the electron fluid's compressibility. The value of this constant 
depends on the frequency regime one is interested in. We have $\beta_{\rm H}\sim v_{F}/\sqrt{3}$ 
is appropriate at low frequency, while $\beta_{\rm H}\sim v_{F}\sqrt{3/5}$ should be used in case 
of a high frequency dynamics  \cite{Bloch34,Barton79}. Here we chose this second value since we will 
be interested in phenomena around the plasma frequency. 
From the above equation, we can directly infer the longitudinal and transverse components of the 
dielectric tensor. If we follow the above-stated premise that the ionic background does not provide 
additional contributions to the dielectric behavior from bound charges, we obtain
\begin{subequations}
\begin{align}
   \epsilon_{t}(\omega)
   &=
   \epsilon_{\rm D}(\omega)
   =
   1-\frac{\omega^{2}_{p}}{\omega(\omega+\imath \Gamma)}, \\
   \epsilon_{l}(K,\omega)
   &=
   1-\frac{\omega^{2}_{p}}{\omega(\omega+\imath \Gamma)-\beta_{\rm H}^{2}K^{2}},
\end{align}
\label{simpleNL}
\end{subequations}
where, the transverse dielectric constant $\epsilon_{t}(\omega)$ is identical to the standard (spatially local) 
Drude dielectric constant $\epsilon_{\rm D}(\omega)$. Thus, in the hydrodynamic model, only the longitudinal 
part of the electric field is actually affected by the nonlocal behavior of the metal.

Following Ref. \cite{Mochan87} (see also \cite{Yan12}), inside the nonlocal medium, the field is no longer 
transverse and is instead given by a superposition of left- and right-propagating waves with transverse and 
longitudinal wave vectors, $k_{\rm D}$ and $k_{\rm B}$, respectively, where
\begin{subequations}
\begin{align}
   k_{\rm D}
   &=
   \sqrt{\frac{\omega^{2}}{c^{2}}\epsilon_{\rm D}(\omega)-k^{2}}\\
   k_{\rm B}
   &= 
   \sqrt{\frac{\omega^{2}}{c^{2}}\frac{\epsilon_{\rm D}(\omega)}{\chi(\omega)}-k^{2}},
   \quad \text{with} \quad
   \chi(\omega)=\frac{\omega}{\omega+\imath \Gamma}\frac{\beta_{\rm H}^{2}}{c^{2}}.
\end{align}
\end{subequations}
The wave vector of the longitudinal wave fulfills $\epsilon_{l} \left( \sqrt{k^{2}+k^{}_{\rm B}},\omega \right)=0$.\par
The continuity of the tangential components of $\mathbf{E}$ and $\mathbf{B}$ relates two unknown coefficients 
(per polarization) in the dielectrics with four unknowns (two transverse and two longitudinal) inside the 
metal. This requires two additional boundary conditions (ABCs) in order to arrive at a well-determined system 
of equations. In case of the hydrodynamic model, it is reasonable to assume that the density of free carriers in 
the metal does not create any singularity at the dielectrics/metal interface. An application of Gauss' theorem 
immediately yields that the normal component of the displacement field $\mathbf{D}$ must be continuous 
across the interface
\begin{equation}
   \epsilon(\omega) E_{z,{\rm lc}} = E_{z,{\rm m}},
\end{equation}
where, $E_{z,{\rm m}}$ and $E_{z,{\rm lc}}$, respectively, denote the electric field in the metal and in the
(spatially local) dielectric material. In other words, the normal component of the electric fields exhibits 
a jump across the metal-dielectric interface, the magnitude of which is governed by the value of $\epsilon(\omega)$.
Using the continuity equation of the electric charge, the above relation of the normal component of the electric
field is also equivalent to the continuity of the orthogonal component of the current of free carriers across 
the interface. If in the dielectric there are no free carriers, this is equivalent to a vanishing $j_{z}$ at 
the interface of the metal. A second boundary condition that ensures consistent optical properties is obtained 
by requiring the scalar electric potential $\phi$ to be continuous across an interface \cite{Mochan87,Ruppin84}. 
With these two ABCs and following a reasoning similar to the one described in the previous subsection, we have  
the $p$-polarization
\begin{equation}
   \begin{pmatrix}
      E\\
      c B\\
      E_{z}\\
      \phi
   \end{pmatrix}^{p}_{z=\frac{p_{1}}{2}+p2}
   =
   \mathbb{Y}(p_{2})
   \begin{pmatrix}
      E\\
      c B\\
      E_{z}\\
      \phi
   \end{pmatrix}^{p}_{z=\frac{p_{1}}{2}}.
\label{Hydro4}
\end{equation}
Here, the transfer matrix $\mathbb{Y}(p_{2})=\mathbb{Z}^{p}_{\rm H}\mathbb{P}_{\rm H}(p_{2})[\mathbb{Z}^{p}_{\rm H}]^{-1}$ 
may be computed from the interface matrix \cite{Mochan87}
\begin{equation}
   \mathbb{Z}^{p}_{\rm H}
   =
   -\frac{Z_{0}}{2}
   \begin{pmatrix}
      Z^{p}_{\rm D} & Z^{p}_{\rm D} &         \imath k &         \imath k \\
                  1 &            -1 &                0 &                 0\\
         -W_{\rm D} &     W_{\rm D} & \imath k_{\rm B} & -\imath k_{\rm B}\\
                  0 &             0 &               -1 &                -1
\end{pmatrix},
\end{equation}
and the propagation matrix 
\begin{equation}
   \mathbb{P}_{\rm H}(z)=\mathrm{diag}\{e^{\imath k_{\rm D}z} ,e^{-\imath k_{\rm D}z},e^{\imath k_{\rm B}z},e^{-\imath k_{\rm B}z}\},
\end{equation} 
where we have introduced the abbreviations
\begin{equation}
   Z_{\rm D}^{p} = \frac{c k_{\rm D}}{\omega\epsilon_{\rm D}(\omega)}, 
   \quad 
   Z_{\rm D}^{s}=\frac{\omega}{c k_{\rm D}}, 
   \quad 
   W_{\rm D}=\frac{c k}{\omega\epsilon_{D}(\omega)}.
\end{equation}
The main difference of these ABCs for the hydrodynamic model with respect to the local case is that
in order to take into account the longitudinal waves (bulk plasmons) in the metals we have to add the 
extra degrees of freedom, $E_{z}$ and the potential $\phi$, in the description of the field. These
longitudinal waves (bulk plasmons) can only be excited when the electric field in the dielectric 
exhibits a non-zero component orthogonal to the surface. As this is not the case for s-polarized
radiation, we can take over the results of the description of s-polarized waves in local (metallic) 
media
\begin{equation}
   \begin{pmatrix}
      E \\
      c B
    \end{pmatrix}_{z=\frac{p_{1}}{2}+p_{2}}^{s}
    =
    \mathbb{M}^{s}_{\rm H}(p_{2})
    \begin{pmatrix}
       E \\
       c B
    \end{pmatrix}_{z=\frac{p_{1}}{2}}^{s},
\end{equation}
where the transfer matrix $\mathbb{M}^{s}_{\rm H}\left(p_{1}\right)$ is
\begin{equation}
   \mathbb{M}^{s}_{\rm H}\left(p_{2}\right)
   =
   \begin{pmatrix}
      \cos(k_{\rm D}p_{2})                             & -\imath \sin(k_{\rm D}p_{2}) Z^{s}_{\rm D}\\
      -\imath \sin(k_{\rm D}p_{2})/Z^{\sigma}_{\rm D} & \cos(k_{\rm D}p_{2})
   \end{pmatrix}
\end{equation}
The situation is entirely different in the case of p-polarized radiation: In the local dielectric material
just next to the interface, we have 
\begin{equation}
   E^{p}_{z,\rm lc} = -c \,\frac{W_{0}}{\epsilon(\omega)}B_{\rm lc}^{p}
   \quad \text{where}\quad 
   W_{0} = \frac{c k}{\omega}.
\end{equation}
The boundary condition at the interface implies that $ \epsilon(\omega) E_{z,{\rm lc}} = E_{z,{\rm m}}$ and 
$B_{\rm lc}^{p} = B_{\rm m}^{p}$, leading to the relation $E^{p}_{z,\rm m} = -c \,W_{0}B_{\rm m}^{p}$ which
is valid inside the metal just next to the interface \cite{Yan12,Mochan87}. This allows us to eliminate $E_{z}$ 
and $\phi$ from Eq.\eqref{Hydro4} so that we finally obtain
\begin{equation}
   \begin{pmatrix}
      E \\
      c B
   \end{pmatrix}_{z=\frac{p_{1}}{2}+p2}^{p}
   =
   \mathbb{M}^{p}_{\rm H}(p_{2})
   \begin{pmatrix}
      E\\
      c B
    \end{pmatrix}_{z=\frac{p_{1}}{2}+p2}^{p}.
\end{equation}
The entries of the transfer matrix (see eq. \eqref{MnlMatrix}) in the hydrodynamic model for p-polarization as thus given by
\begin{subequations}
   \begin{equation}
      \mathrm{cs}^{p}_{\rm nl}(p_{2})
      =
      \mathbb{Y}_{11} - \mathbb{Y}_{14} \frac{W_{0} \mathbb{Y}_{21}+\mathbb{Y}_{31}}{W_{0}\mathbb{Y}_{24} + \mathbb{Y}_{34}} ~,
   \end{equation}
   \begin{multline}
      \imath \mathcal{Z}^{p}_{\rm right}(p_{2})
      =
      \mathbb{Y}_{12}-W_{0}\mathbb{Y}_{13}\\
      -\mathbb{Y}_{14}
      \left(          \frac{W_{0}\mathbb{Y}_{22}+\mathbb{Y}_{32}}{W_{0}\mathbb{Y}_{24}+\mathbb{Y}_{34}} - 
            W_{0}\frac{W_{0}\mathbb{Y}_{23}+\mathbb{Y}_{33}}{W_{0}\mathbb{Y}_{24}+\mathbb{Y}_{34}}
      \right) ~,
   \end{multline}
   \begin{equation}
      \frac{\imath}{\mathcal{Z}^{p}_{\rm left}(p_{2})}
      = 
      \mathbb{Y}_{21}-\mathbb{Y}_{24} \frac{W_{0} \mathbb{Y}_{21}+\mathbb{Y}_{31}}{W_{0}\mathbb{Y}_{24}+\mathbb{Y}_{34}} ~.
   \end{equation}
\end{subequations}

\begin{figure*}
\center
\includegraphics[width=8cm]{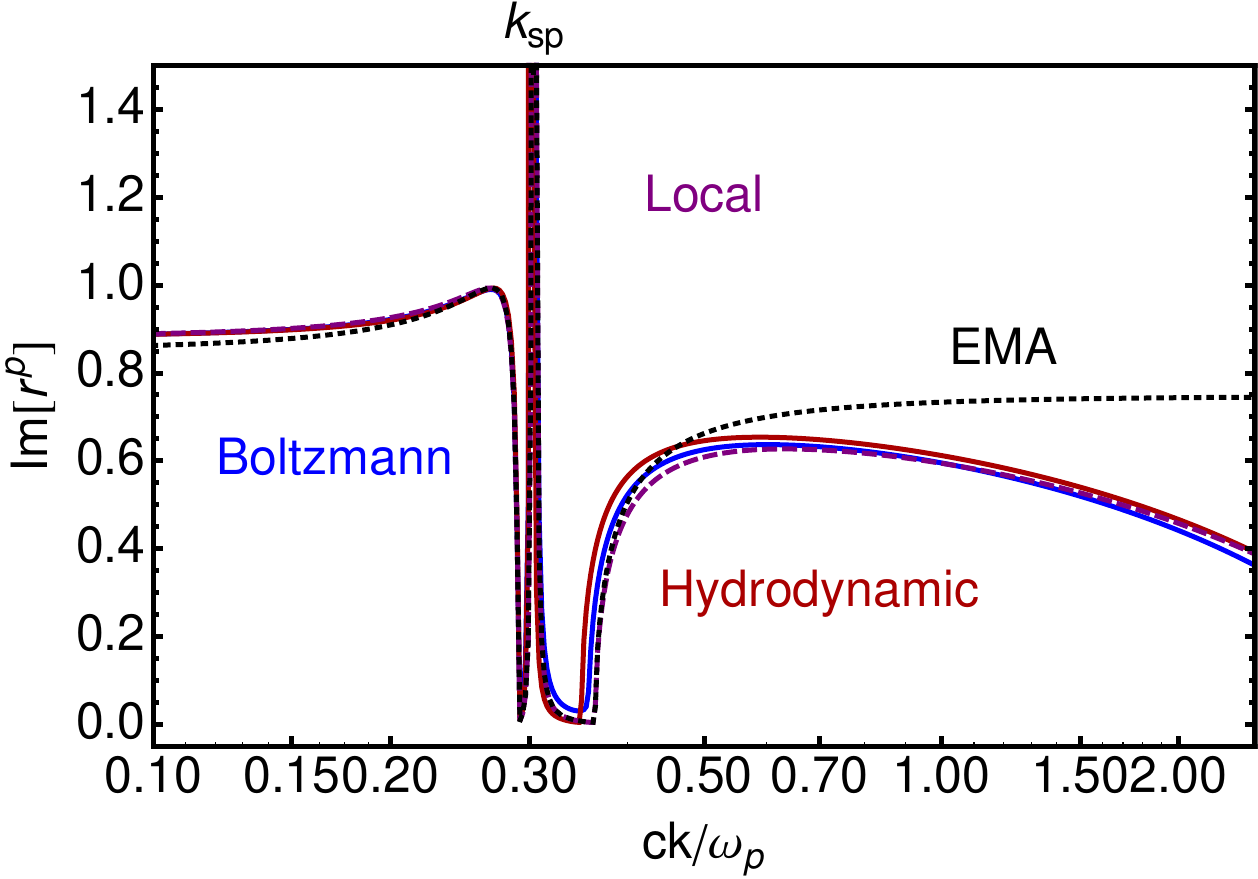}
\hspace{2mm}
\includegraphics[width=8.3cm]{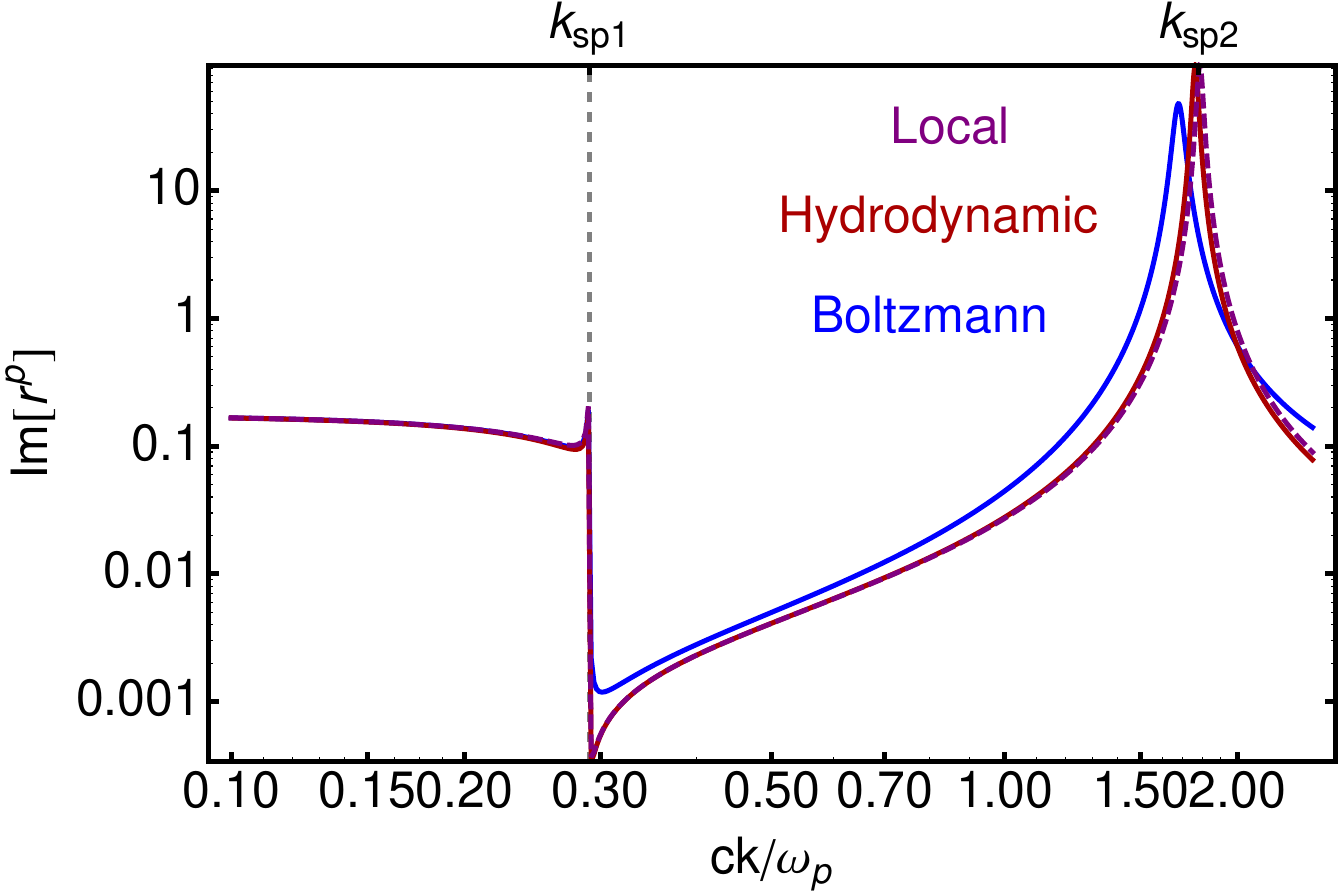}
\includegraphics[width=8.3cm]{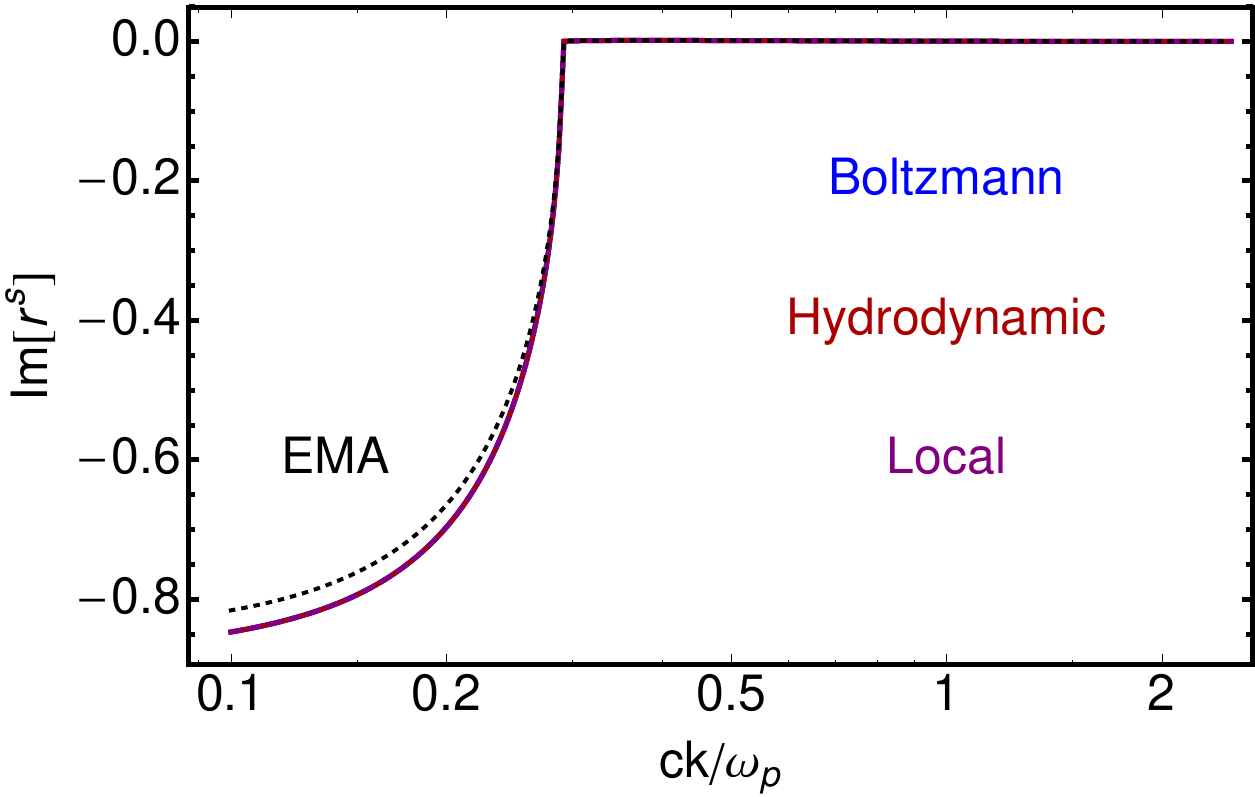}
\hspace{2mm}
\includegraphics[width=8.3cm]{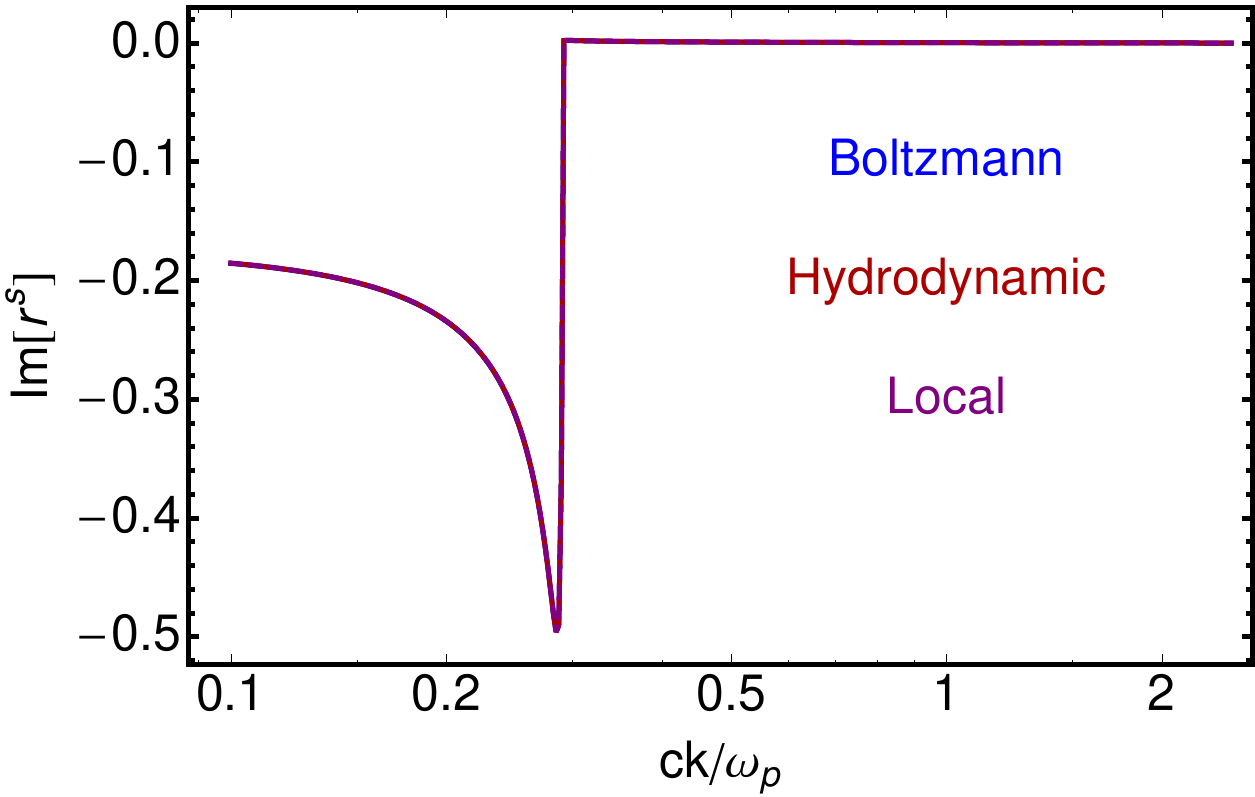}
\caption{(Color online) 
         Top row: Lateral wave-vector dependence of the imaginary part of the reflection coefficients for 
				          p-polarized light for a infinite number of silver/silica bilayers (left panel) and a thin 
									silver layer on top of a silica half-space (right panel). 
									The frequency is fixed to $\omega=0.2 \omega_{p}$ and the thickness of the silica and
									silver layers are $p_1 = 0.2 c/\omega_{p}\sim 4.4$nm and $p_2= 0.1 c/\omega_{p}\sim 2.2$nm,
									respectively (see Fig. \ref{ModulatedRegionHH}). 
			Bottom row: Same as the top row but for s-polarized light. 				
		  In each plot, the different curves correspond to different material models for silver: The nonlocal SCIB 
			model based on the Boltzmann equation (blue solid line), the local Drude model (purple dashed line) and 
			the nonlocal hydrodynamic model (red solid line).  
			For comparisons with the case of the periodic structure, also the predictions of the effective medium 
			approximation (EMA, black dashed) \cite{Kidwai12} are depicted.
\label{ImRps}}
\end{figure*}

\subsection{Discussion of the Nonlocal Material Models}
The two material models described above, display several analogies but also profound differences 
\cite{Feibelman82,Ford84}. 
If we consider the expressions in Eq.\eqref{etabeta}, the residue theorem allows us to show that $\eta^{p}(z)$ 
can be written as the sum of waves that propagate with wave vectors that are solutions of 
$K^{2}-\epsilon_{t}(K,\omega)\omega^{2}/c^{2}=0$ and $\epsilon_{t}(K,\omega)=0$. Clearly, these solutions 
correspond to transverse and longitudinal waves, respectively. This correspondence between the models is, 
however, only qualitative as the expressions for longitudinal and transverse dielectric functions are quite 
different. For instance, while the hydrodynamic model only exhibits nonlocal modifications to the longitudinal 
part of the electromagnetic field, the SCIB predicts nonlocal modifications for both the longitudinal and the 
transverse  part of the field.  
Probably the most apparent difference between the two models concerns the boundary conditions at the interface. 
While the SCIB model relies on the symmetries of the Boltzmann equation to determine the behavior of electrons
at an interface with dielectric materials, the hydrodynamic description uses the non-locality to implement a 
finite density of electrons at the interface which removes the discontinuity in the $z$-component of the 
displacement field. 

\begin{figure*}
\includegraphics[width=8.2cm]{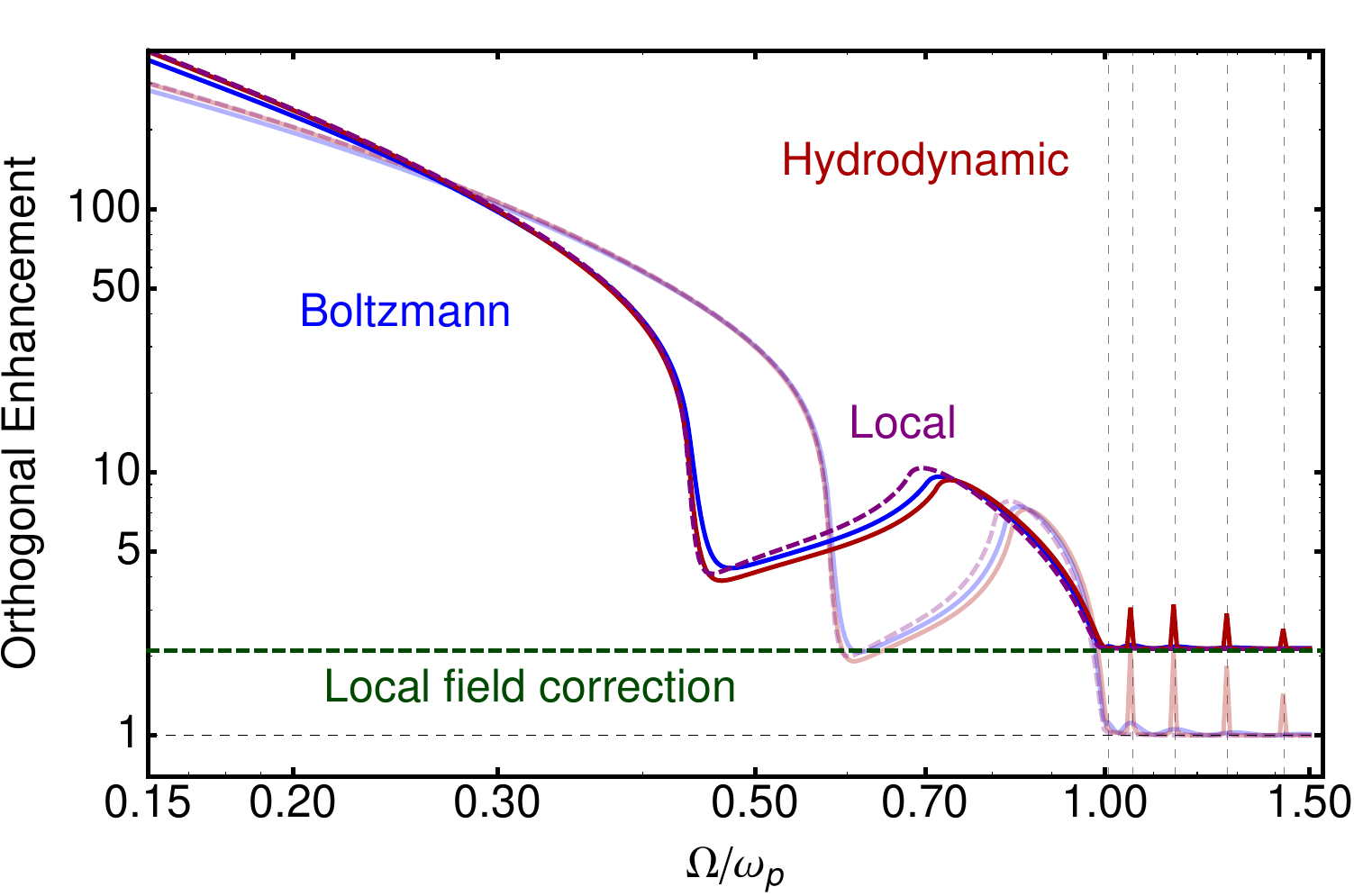}
\includegraphics[width=8.2cm]{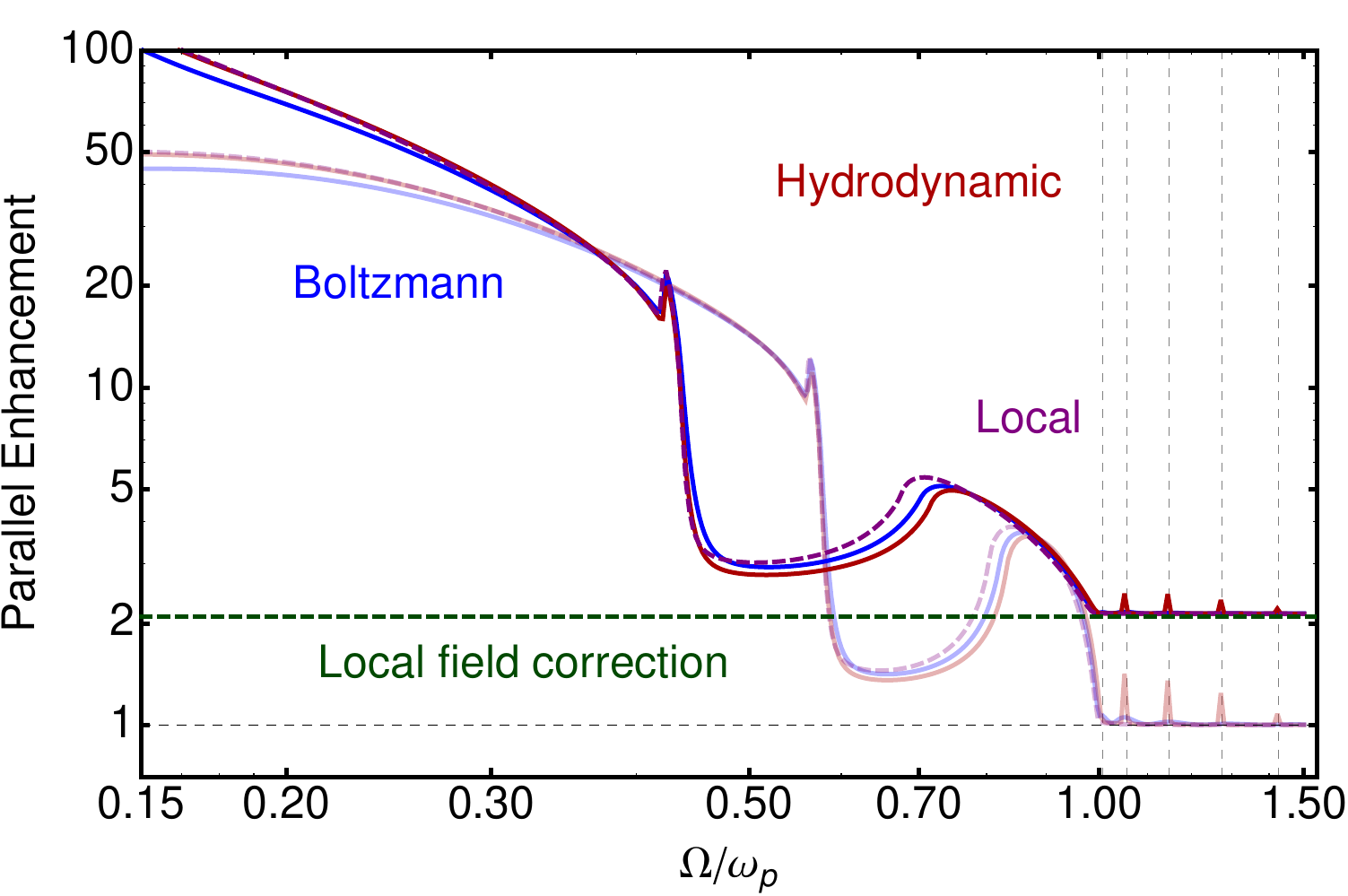}
\caption{(Color online) Spontaneous emission enhancement factor derived from the orthogonal (left) and parallel 
         (right) components of the scattering Green's tensor of an infinite sequence of alternating silver and 
				 silica layers. The different curves correspond to different material models for silver: The nonlocal 
				 SCIB model based on the Boltzmann equation (blue solid line), the local Drude 
         model (purple dashed line) and the nonlocal hydrodynamic model (red solid line).  
         In order to highlight the impact of the local-field corrections, the same computations have been carried 
				 out for the same parameters except for replacing silica by vacuum. The results are displayed in the 
				 semi-transparent curves. In particular, the incorporation of local-field effects leads to a significant 
				 broadening of the plasmon resonance. The gray dashed lines indicate the position of the odd bulk plasmon 
				 resonance as give in eq.\eqref{Bulkres} ($\omega^{\rm B}_{2n+1}$).
         See the text for details regarding the geometric and material parameters.
\label{OrthPar}}
\end{figure*}

Despite these differences, the two non-local models provide qualitatively similar results for the reflection 
coefficients of the infinitely periodic structures and of a thin silver slab embedded in silica matrix. In 
Fig. \ref{ImRps} we represent the imaginary part of the corresponding reflection coefficients for the p- and 
s-polarization. The quantities are plotted as functions of the lateral wave-vector for a fixed frequency 
($\omega=0.2 \omega_{p}$). We note that for the periodic structure the usual effective medium approach (EMA) 
in terms of a local dielectric functions \cite{Orlov11,Chebykin11} provides a good description of the system 
for sufficiently small wave-vectors (see Fig.\ref{ImRps}) \cite{Kidwai12}. As expected, 
\cite{Kidwai12} 
the agreement degrades at larger wave-vectors. The plasmon resonance of the periodic structure ($k_{\rm sp}$, 
see top left panel of Fig. \ref{ImRps}) coincides with the surface-plasmon-polariton (SPP) of a silica/metal 
interface 
\begin{equation}
k_{\rm sp}(\omega)=\frac{\omega}{c}\sqrt{\frac{\epsilon(\omega)\epsilon_{\rm D}(\omega)}{\epsilon(\omega)+\epsilon_{\rm D}(\omega)}}~. 
\end{equation}
The thin single metallic layer exhibits two resonances associated with the symmetric and antisymmetric coupling 
of the SPPs on the two metal/silica interfaces ($k_{\rm sp1}$ and $k_{\rm sp2}$, see top right panel of Fig. \ref{ImRps})
which in other context's are know as the short-range and the long-range SPP, respectively \cite{Berini09}). 
In the local description for frequency smaller than $\omega_{p}$ and small thickness the resonances' positions 
are approximatly given by
\begin{subequations}
\begin{equation}
 k_{\rm sp1}(\omega)\sim \frac{\omega}{c}\sqrt{\epsilon(\omega)} ~,
\end{equation}  
\begin{equation}
 k_{\rm sp2}(\omega)\sim \sqrt{\left[\frac{2\epsilon(\omega)}{p_{2}\epsilon_{\rm D}(\omega)}\right]^{2}+\frac{\omega^{2}}{c^{2}}\epsilon(\omega)} ~.
\end{equation}  
\end{subequations}
which correspond to the values for the symmetric (near the light cone) and anti-symmetric SPPs. The anti-symmetric 
resonance is much stronger than the symmetric resonance. It is also worth noting that for the anti-symmetric SPP
the SCIB gives rise to a value which is different from the value for the local and the hydrodynamic description. 
For s-polarized radiation, the behavior is much simpler and the  nonlocality only slightly affects the reflection 
coefficients. In all cases, a characteristic abrupt change occurs at the light cone, i.e. for 
$k\sim \omega\sqrt{\epsilon(\omega)}/c$.

\section{Results}
\label{results}

We now apply the above formalism to study the modified radiation dynamics of an emitter embedded in two distinct
structures. The first structure consists of a central cavity silica layer ($D =2 c/\omega_{p}$) that is 
symmetrically sandwiched between infinite sequences of bilayers of silver ($p_2 = 0.1 c/\omega_{p}$) 
and silica ($p_1= 0.2 c/\omega_{p}$) as depicted in Fig. \ref{ModulatedRegionHH}. The second structure 
comprises the same central cavity silica layer that is symmetrically sandwiched between two silver 
layers with thickness $p_2 =  0.1 c/\omega_{p}$ and this composite slab-structure is embedded into two half spaces of silica. 
The dielectric properties of silica are described via a three-oscillator model
\cite{Gunde00} 
and we consider the above-discussed and widely used material models for silver, i.e., the local Drude model, 
the SCIB model and the hydrodynamic model. All these models employ the same plasma frequency $\omega_{p}=8.89$ eV ($c/\omega_{p}\sim 22$nm) 
and damping constant $\Gamma=0.018$ eV. Additionally, the SCIB model and the hydrodynamic model use the Fermi 
velocity $v_{F} =1.39 \times 10^{6}$ m/s) of silver
\cite{Iorsh12}.
In both of the above structures, we position an emitter midway in the cavity layer ($d = c/\omega_{p}$) and the radius
of the real-cavity model for the local field correction is $R = 10^{-2} c/\omega_{p}$.

\subsection{Decay Enhancement and Frequency Shift in Infinite Periodic Structure}

\begin{figure}[h]
\includegraphics[width=8.3cm]{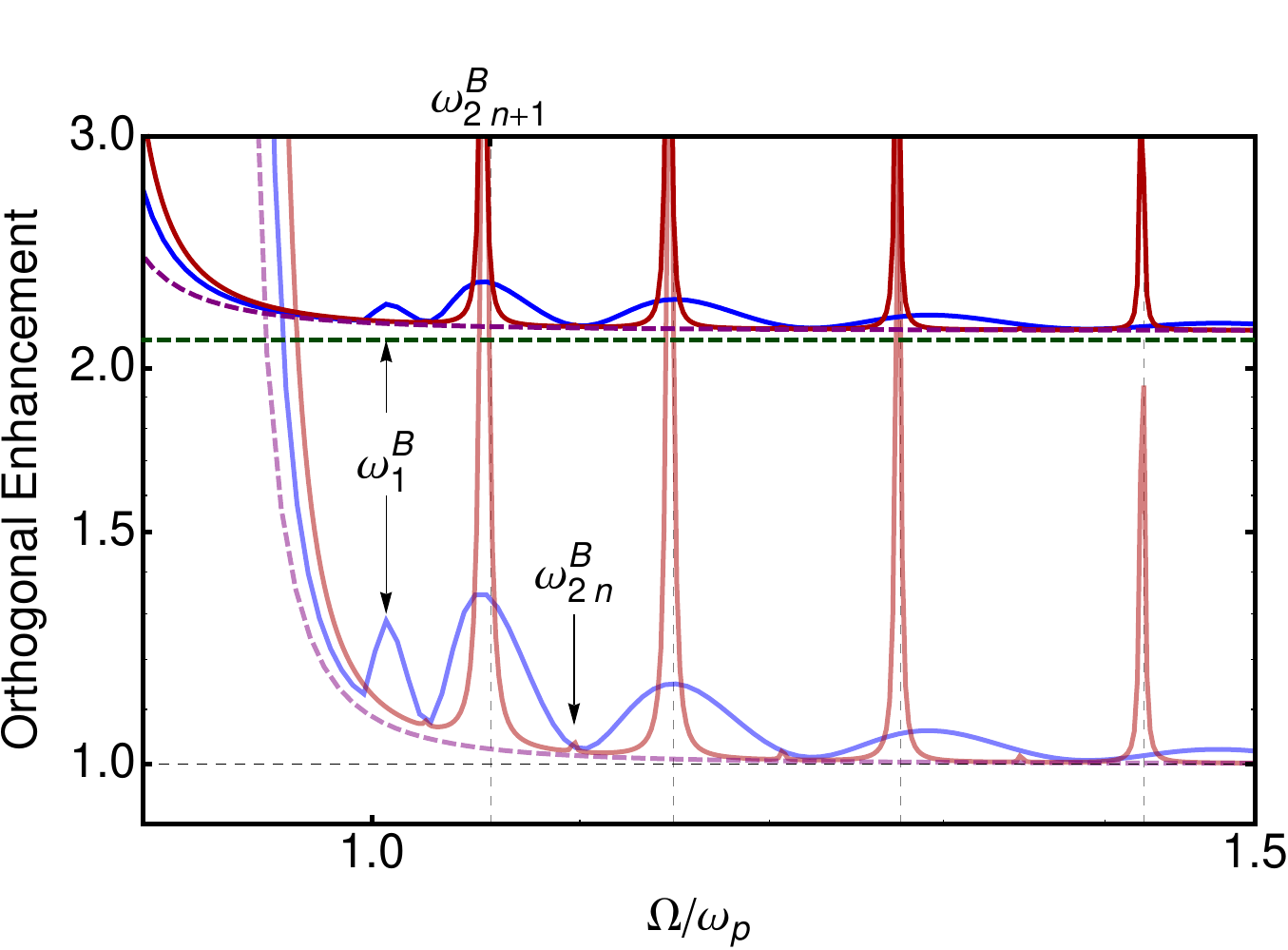}
\caption{(Color online) Bulk plasmon resonances in the orthogonal enhancement of the spontaneous 
          decay of an emitter located in a cavity formed by two thin metallic layers. The position 
					and behavior of these resonances is similar in all configurations considered in this work. 
					The different curves correspond to different material models for silver: The local Drude 
					model (purple dashed line), the nonlocal hydrodynamic model (red solid line) and the SCIB 
					model based on the Boltzmann equation (blue solid line).  
					See the text for details regarding the geometric and material parameters.
          In order to highlight the impact of the local-field corrections, the same computations 
					have been carried out for the same parameters except for replacing silica by vacuum. The 
					results are depicted in the semi-transparent curves.  
\label{BulkOrtSlab}}
\end{figure}

In Fig. \ref{OrthPar} we depict the results of the spontaneous emission enhancement 
factor for a dipole oriented orthogonal and parallel to the stacking direction of the above-described infinite 
periodic structure. For comparison, we have also included the results of computations where silica has been 
replaced by vacuum (or air). As expected, the nonlocal material models lead to a slight blue shift of the main 
plasmon resonance around $\omega/\omega_p =1\sqrt{2}\sim 0.7$ relative to the local Drude model. The decay rates 
essentially follow the dispersion relation of the surface plasmons coupled across the cavity containing the 
emitter. From the expressions of the orthogonal and parallel components of the Green's tensor in eqs.\eqref{greenT} 
one deduces that the orthogonal enhancement is associated with the dispersion relation of the anti-symmetric 
cavity surface plasmon while the behavior of the parallel enhancement can be associated with the dispersion 
relation of the symmetric cavity surface plasmon \cite{Mochan87,Haakh13,Intravaia05,Intravaia07}: The integrals 
in eqs.\eqref{greenT} can be approximatively evaluated as the residues for the corresponding cavity plasmon. 
Upon using dimension less variables $\omega/c\to \omega d/c $ and $k\to k d$ in eqs.\eqref{greenT}, one can infer that the behavior at low frequencies is equivalent to a reduction of the distance between the emitter and the interface explaining the large enhancement of the decay. In this same region ($\omega/\omega_{p}\lesssim 0.2$) the hydrodynamic model gives results very similar to the local description while the SCIB model produces slightly a different prediction (more pronounced when we use silica instead of vacuum for the dielectric layer).
At frequencies higher than the plasma frequency, we observe additional resonances for the nonlocal materials 
models. These resonances corresponds to the excitation of bulk plasmons which are known to appear in nonlocal 
descriptions of the metal beyond the plasma frequency \cite{Mochan87}. It is worth noting that while these bulk 
plasmon resonances occur at roughly the same positions for the SCIB and the hydrodynamic model (a shift appears 
at large frequencies) they are much less pronounced and wider for the SCIB model (see Fig.\ref{BulkOrtSlab}). 
Approximately, the bulk plasmon resonances are given by
\begin{equation}
\omega^{\rm B}_{n}\approx\sqrt{\omega_{p}^{2}+\left(n\frac{\beta\pi}{p_{2}}\right)^{2}}.
\label{Bulkres}
\end{equation}

In the hydrodynamic description only the odd frequencies couple to the external radiation \cite{Mochan87} while the 
even resonances are almost decoupled from the external field and can be excited only minimally (see Fig.\ref{BulkOrtSlab}). 
We observe a similar behavior for the SCIB model with the exception of the lowest bulk plasmon frequency $\omega^{\rm B}_{1}$: 
In the hydrodynamic model this resonance lies in a band gap which forbids any propagation \cite{Mochan87}. The situation is 
different for the model based on the Boltzmann equation, where we clearly observe a resonance at $\omega^{\rm B}_{1}$. 

\begin{figure}[hh]
\includegraphics[width=8.3cm]{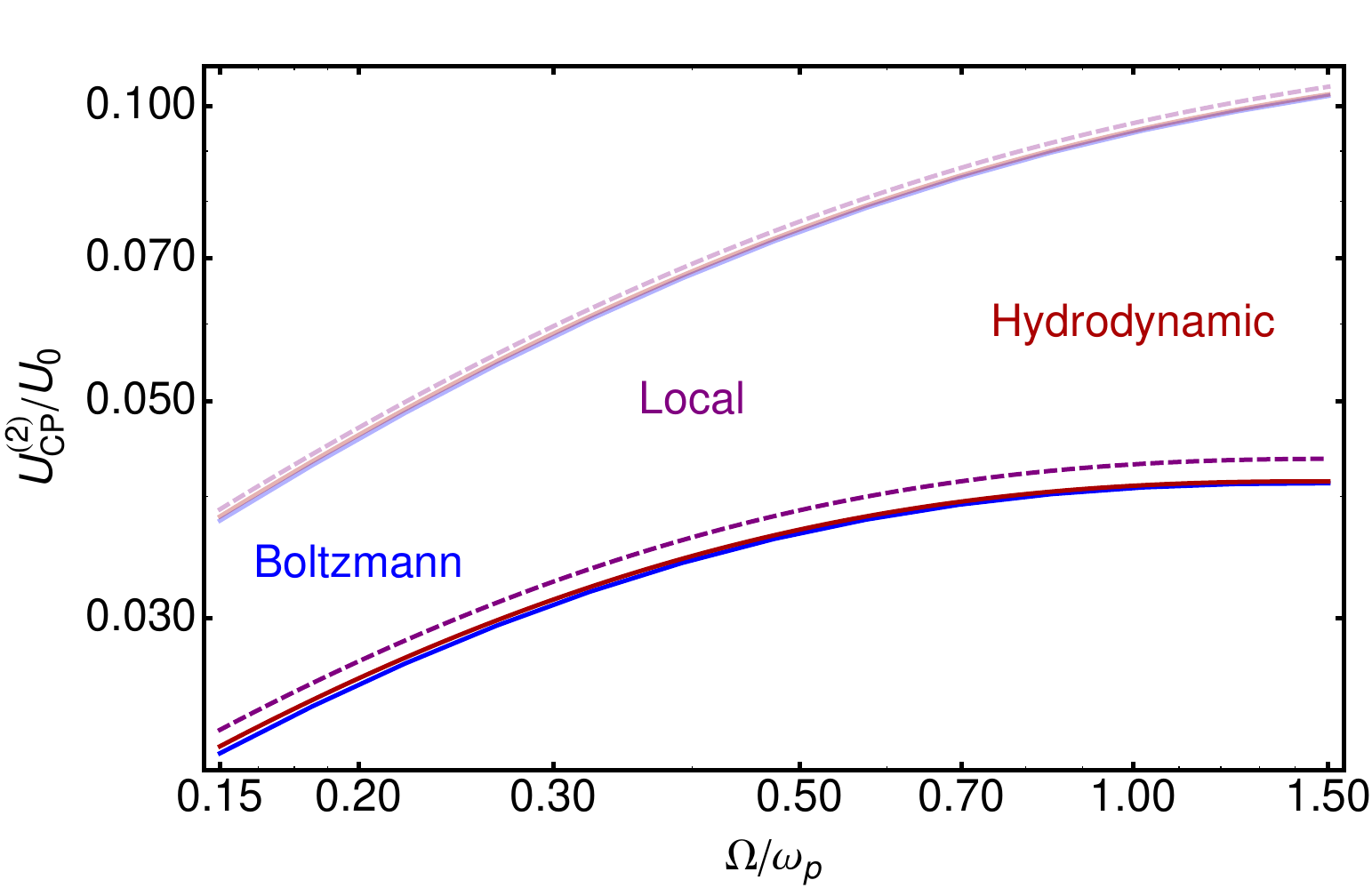}
\caption{(Color online) Geometric frequency shift for an emitter in a silica nano-cavity sandwiched between
         infinite sequences of alternating silver and silica layers. The parameters and the models are 
				 the same as those used for describing the spontaneous decay (see main text). The values are 
				 normalized to $U_{0}=-\hbar \omega_{p} (\alpha_{g} \omega_{p}^{3}/c^{3})(2\pi\epsilon_{0})^{-1}$ 
				 and depicted as a functions of the emitter's transition frequency. 
				 The different curves correspond to different material models for silver: The local Drude model 
				 (purple dashed line), the nonlocal hydrodynamic model (red solid line) and the SCIB model based 
				 on the Boltzmann equation (blue solid line). 
				 See the text for details regarding the geometric and material parameters.
				 In order to highlight the impact of the local-field corrections, the same computations have been 
				 carried out for the same parameters except for replacing silica by vacuum. The results are depicted
				 in the semi-transparent curves.  
         The local-field corrections significantly reduce the magnitude of these geometry-induced shifts 
				 relative to vacuum.      
\label{FreqShift}}
\end{figure}

The differences between the nonlocal models are directly connected to the different ways the wave vector (non-locality) 
and dissipation enter in eqs.\eqref{epsilonSCIB} and \eqref{boltzmann} with respect to eqs.\eqref{simpleNL}  
and, clearly, also to the different boundary conditions discussed in section \ref{nonlocal}.
Therefore, experimental studies on the spontaneous emission enhancement in such systems for frequencies above the plasma 
frequency may be able to probe the nature of the plasmonic material.

Upon comparing the results for vacuum with those of silica as the dielectric material, the offset originating from the 
local-field corrections and the red-shift of the curves as well as of the SPP resonance is clearly visible. 
Again, this (expected) behavior can be understood in connection with the dispersion relation of the symmetric and anti-symmetric cavity surface plasmons which are expected to red-shift in presence of the dielectrics.
Furthermore, the local-field 
corrections lead to a rather significant broadening of the main plasmon resonance characteristics. Conversely, the positions 
of the bulk plasmon resonances are not affected by the local field correction and there hardly is any additional broadening 
for both models. Instead, we observe a reduction in the peak hight.

\begin{figure*}[hht]
\includegraphics[width=8.4cm]{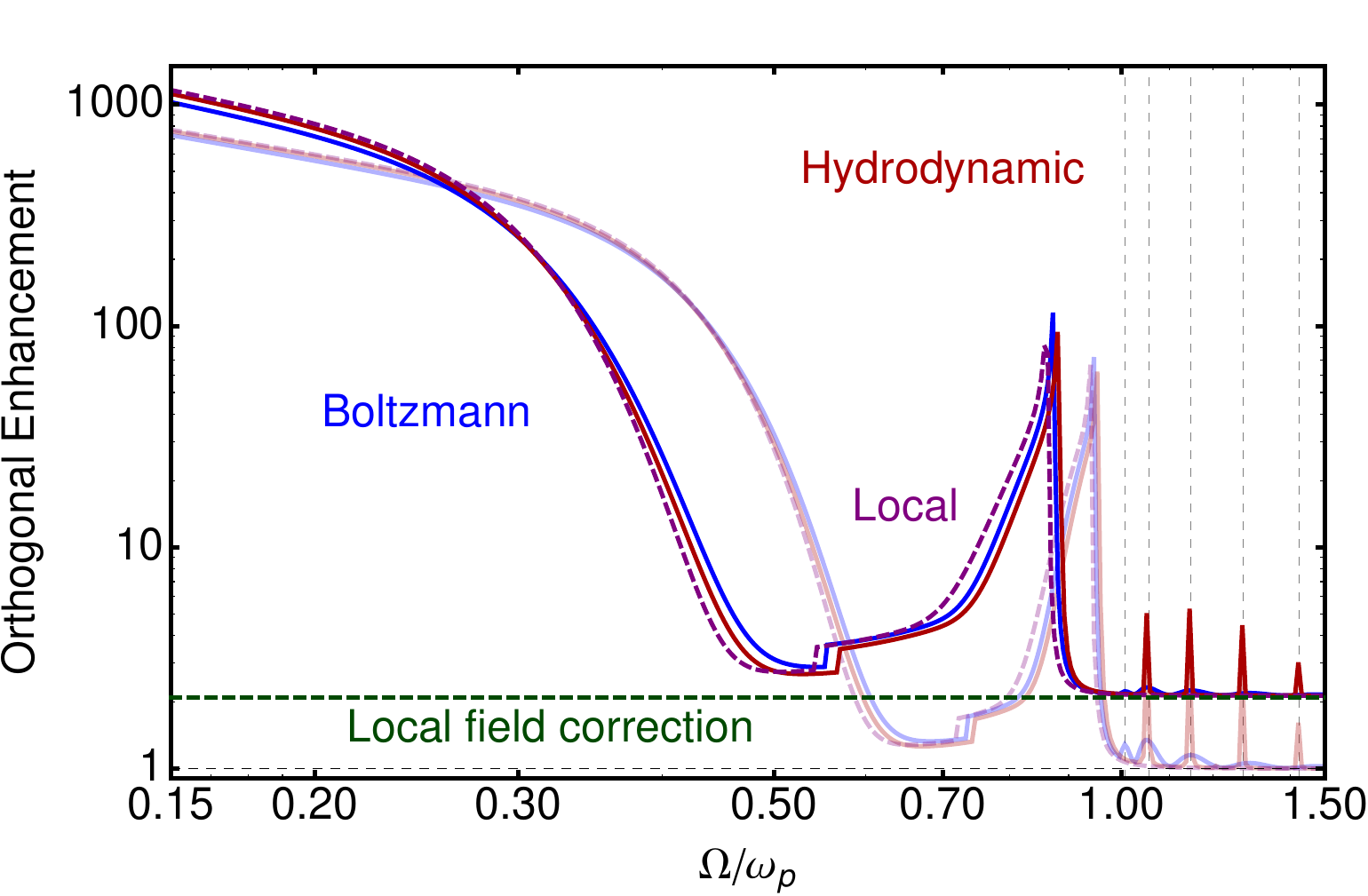}
\includegraphics[width=8.3cm]{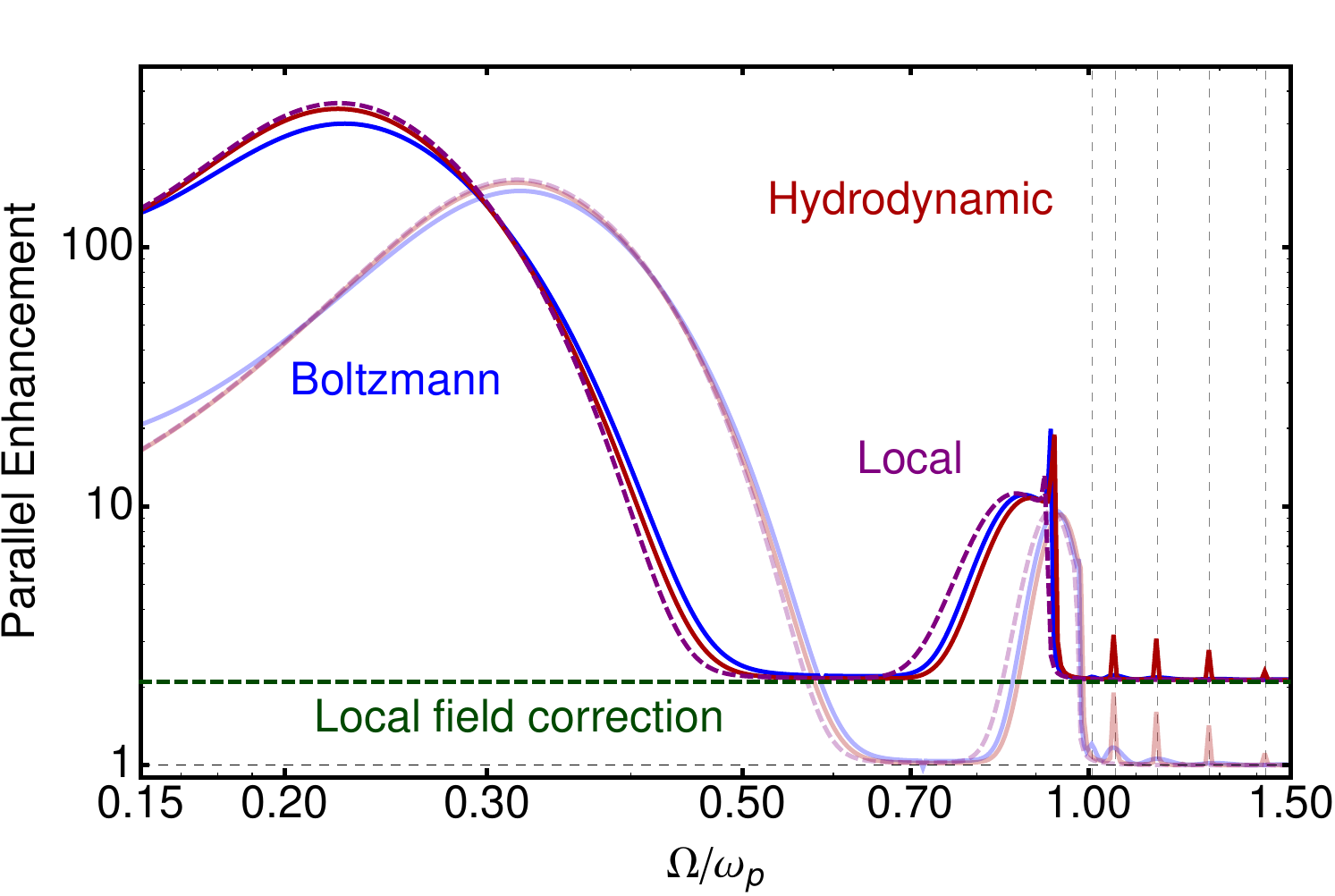}
\caption{(Color online) Spontaneous emission enhancement factor derived from the orthogonal (left) and parallel (right) 
         components of the scattering Green's tensor of composite slab structure consisting of a central silica layer 
				 sandwiched between two silver layers and completely embedded in silica. The emitter is located in the center 
				 of central silica layer. The different curves correspond to different material models for silver: The nonlocal 
				 SCIB model based on the Boltzmann equation (blue solid line), the local Drude model (purple dashed line) and 
				 the nonlocal hydrodynamic model (purple solid line).  
         In order to highlight the impact of the local-field corrections, the same computations have been carried out 
				 for the same parameters except for replacing silica by vacuum. The results are displayed in the semi-transparent 
				 curves. 
         The incorporation of local-field effects leads to a significant broadening of the plasmon resonance. 
				 The gray dashed lines indicate the position of the odd bulk plasmon resonance as give in eq.\eqref{Bulkres} 
				 ($\omega^{\rm B}_{2n+1}$).
         See the text for details regarding the geometric and material parameters.
\label{OrthParSlab}}
\end{figure*}

In addition, in Fig. \ref{FreqShift} we display the results for the geometrical frequency shift experienced 
by the emitter in the above-discussed infinite structure. Owing to the fact that this shift results from an integration 
over imaginary frequencies (c.f. Eq. \eqref{geoshift}), no characteristic features resulting from plasmon resonances 
are visible and the results for the local Drude model and the nonlocal models are very similar and this is in agreement 
with previous works \cite{Esquivel03,Esquivel04,Roman-Velazquez04,Esquivel-Sirvent05,Svetovoy05}. Also, the differences 
between the SCIB and the hydrodynamic description are less prominent and both models essentially provide the same result.  
Nevertheless, the local-field corrected computations for silica yield geometry-induced shifts that are significantly 
reduced with respect to the computations for vacuum. This is the result of an effective screening provided by the dielectric 
material.

\subsection{Decay Enhancement for a Finite Slab Structure}

In Fig. \ref{OrthParSlab}, we display the spontaneous emission enhancement factor for a dipole oriented 
orthogonal and parallel to the stacking direction of the above-described slab structure. For comparison, we have 
included the results of computations where silica has been replaced by vacuum (or air).
Also in this case the enhancement factor shows the features described above.
As before, we observe the characteristic blue-shift of the main plasmon resonance between the local Drude description 
and the nonlocal descriptions for the plasmonic layers as well as the occurence of bulk plasmon resonances for 
frequencies above the plasma frequency, strong resonances for the hydrodynamic model and weak resonances for the 
SCIB model. 
Similarly, the comparison of the local-field corrected computations for silica with the computations for vacuum 
reveal that in the case of silica the main plasmon resonance is much broader and the bulk plasmon resonance peaks 
are suppressed. However, we would like to note that the enhancement values generally are much larger than for the 
infinite system discussed above.
This is for our specific choice of geometric parameters the resonance in the reflection coefficient for the 
thin slab are much stronger as compared to the infinitely layered system (see Fig. \ref{ImRps}).
Finally, in the composite silica-silver slab system for frequencies $\omega/\omega_{p}\lesssim 0.2$, we observe again differences between the emission enhancements related to the nonlocal SCIB description and to the hydrodynamic model. This occurs for both dipole's orientation whereas for these frequencies, the results of the hydrodynamic model agrees rather well with those of the Drude model. In this case we can connect this behavior with the features of the p-reflection coefficient in Fig. \ref{ImRps}.

In summary, we have developed a comprehensive framework for computing decay enhancements and
level shifts for emitters embedded in arbitrary layered structures. This framework is capable 
of including local-field corrections in (weakly absorbing) dielectric systems and can treat
the nonlocal optical properties of metals. 
All these features influence the emitters's dynamic in a non-additive way, which is also amplified by the relative complexity of the surrounding structure. Nevertheless we were able to show that the local-field corrections generally introduce an offset in the spontaneous decay rates and lead to broadening of plasmon resonances below 
the plasma frequency. In addition, local-field corrections effectively reduce geometry-induced 
level shifts. Furthermore, we have found that the differences between the different material
models for metals can be analyzed either by changing the dielectric material between the 
metal layers or by carefully inspecting the decay rates for frequencies above the plasma
frequency. While this may be unrealizable for silver-based structures, we would like to point
out that recent advances in highly-doped semiconductors place their plasma frequency in the
near infrared  \cite{Sadofev13}, thus rendering such investigations experimentally feasible.

\section{Acknowledgments}
We acknowledge support by the Deutsche Forschungsgemeinschaft (DFG) through the sub-projects 
B10 within the Collaborative Research Center (CRC) 951 Hybrid Inorganic/Organic Systems for 
Opto-Electronics (HIOS). 
FI further acknowledges financial support from the European Union Marie Curie People program 
through the Career Integration Grant No. 631571 and through the German-Israeli Project Cooperation 
(DIP) project ``Quantum Phenomena in Hybrid Systems: Interfacing Engineered Materials and 
Nanostructures with Atomic Systems''.



\end{document}